\documentclass[structabstract]{aa}  
\usepackage[bf]{caption}
\usepackage{graphicx}
\usepackage{subfig}
\usepackage{float}
\usepackage{newclude}
\usepackage{url}
\usepackage{multirow}
\usepackage{enumerate}
\usepackage{txfonts}
\usepackage{aalongtable}
\usepackage{natbib}
\bibpunct{(}{)}{;}{a}{}{,} 
\begin{document}
\newcommand{\df}{\mathrm{d.o.f.}}
\newcommand{\thomas}{\citep{Kruehler11}}
\newcommand{\thomast}{\citet{Kruehler11}}
\newcommand{\robert}{(Filgas et al. (2011))}
\newcommand{\sfrunit}{M_{\sun}\,yr^{-1}\,Mpc^{-3}}

   \title{The long $\gamma$-ray burst rate and the correlation with host galaxy properties}
   \author{J.~Elliott\inst{1}
	\and 
	  J.~Greiner\inst{1}
	\and 
	  S.~Khochfar\inst{1}
	\and
	  P.~Schady\inst{1}
	\and 
	  J.~L.~Johnson\inst{1, 2}
	\and
	  A.~Rau\inst{1}
    }

   \institute{
	    Max-Planck-Institut f{\"u}r extraterrestrische Physik, Postfach 1312, Giessenbachstr., 85741 Garching, Germany	 \\
              \email{jonnyelliott@mpe.mpg.de}
           \and 
             Los Alamos National Laboratory, Los Alamos, NM 87545, USA
           }

   \date{Received 02 December 2011 /
	  Accepted 30 January 2012}

	    
  \abstract
   {To answer questions on the start and duration of the epoch of reionisation, periods of galaxy mergers and properties of other cosmological encounters, the cosmic star formation history, $\dot{\rho}_{*}$ or CSFH, is of fundamental importance. Using the association of long-duration gamma-ray bursts (LGRBs) with the death of massive stars and their ultra-luminous nature ($>10^{52}\mathrm{erg\,s^{-1}}$), the CSFH can be probed to higher redshifts than current conventional methods. Unfortunately, no consensus has been reached on the manner in which the LGRBs rate, $\dot{\rho}_{grb}$ or LGRBR, traces the CSFH, leaving many of the questions mentioned mostly unexplored by this method.}
  {Observations by the gamma-ray burst near-infrared detector (GROND) over the past 4 years have, for the first time, acquired highly complete LGRB sample. Driven by these completeness levels and new evidence of LGRBs also occurring in more massive and metal rich galaxies than previously thought, the possible biases of the $\dot{\rho}_{grb}$-$\dot{\rho}_{*}$ connection are investigated over a large range of galaxy properties.}
  {The CSFH is modelled using empirical fits to the galaxy mass function and galaxy star formation rates. Biasing the CSFH by means of metallicity cuts, mass range boundaries, and other unknown redshift dependencies of the form $\dot{\rho}_{grb}\propto\dot{\rho}_{*}(1+z)^{\delta}$, a $\dot{\rho}_{grb}$ is generated and compared to the highly complete GROND LGRB sample.}
  {It is found that there is no strong preference for a metallicity cut or fixed galaxy mass boundaries and that there are no unknown redshift effects ($\delta=0$), in contrast to previous work which suggest values of $Z/\mathrm{Z_{\sun}}\sim0.1-0.3$. From the best-fit models obtained, we predict that  $\sim1.2$\% of the LGRB burst sample exists above $z=6$.}
  {The linear relationship between $\dot{\rho}_{grb}$ and $\dot{\rho}_{*}$ suggested by our results implies that redshift biases present in previous LGRB samples significantly affect the inferred dependencies of LGRBs on their host galaxy properties. Such biases can lead to, for example, an interpretation of metallicity limitations and evolving LGRB luminosity functions.}


   \maketitle

%

\section{Introduction}
\label{sec:Introduction}
Long-duration gamma-ray bursts (LGRBs) are among the most luminous and energetic events to occur in our Universe. First signalled by their prompt high energy emission \citep[observer frame $\gamma$-rays; e.g.,][]{Narayan92}, pinpointing regions of star formation irrespective of host galaxy luminosity, they are then followed by a longer wavelength and longer lasting afterglow \citep[X-ray through to radio; e.g.,][]{Cavallo78, Goodman86, Paczynski86, Meszaros02} from which precise positions, redshifts and host galaxy abundance measurements can be obtained. Such high luminosities allow LGRBs to be detected to high redshifts making them powerful probes of the early Universe. As early as 1993, they were thought to have been the result of the core-collapse of a massive star, under the mechanism named the {\it collapsar model} \citep{Woosley93, Paczynski98, MacFadyen99}. Given strong evidence of an association of a LGRB to a supernova (SN) in 1998, \citep[][SN1998bw]{Galama98} and first conclusive spectroscopic confirmation of SN2003dh associated with GRB 030329 \citep[][]{Stanek03,Matheson03}, the connection of most LGRBs with the core-collapse of massive stars is now unquestionable \citep[for a review, see e.g.,][]{Woosley06b}.

As a result of the collapsar model's connection to the death of massive stars and the small distances that the progenitors travel away from their birth location in their relatively short lives, LGRBs can be used to trace the star formation rate of their host environment. Combined with the fact that LGRBs have been spectroscopically confirmed to be a cosmological phenomena \citep[see e.g.,][]{Metzger97}, they could be used to trace the cosmic star formation history (CSFH; $\dot{\rho}_{*}$). The advantage of using LGRBs rather than conventional methods based on Lyman-break galaxies (LBGs) and Lyman-$\alpha$ emitters (LAEs) \citep[e.g.,][]{Rafelski11}, is that their immense luminosities allow them to be detected up to very high redshifts, e.g., z = 6.7 \citep{Greiner09b}, 8.2 \citep{Tanvir09} and 9.2 \citep[][]{Cucchiara11}. Therefore, the CSFH-LGRB rate (LGRBR) connection has been investigated from as early as 1997 \citep[see e.g.,][]{Totani97, Wijers98}.

Knowing the CSFH to such high redshifts is of fundamental importance for studying galaxy evolution. The current picture of the CSFH~\citep[see e.g.,][]{Madau96, HopkinsBeacom06, Li08, Kistler09} is that there is a steady increase of star formation from $z=0$ to $z=1$ followed by a plateau up to redshift $z\sim4$. However, the shape of the CSFH above $z=4$ remains highly uncertain, where it could continue to plateau \citep{Kistler09}, drop off \citep{Li08} or even increase \citep{Daigne06}.

The achievement of constraining the CSFH to high redshifts would allow many questions about the Universe to be answered, with one of the most sought after being: ``when did reionisation occur?''. To study such questions at high redshift an understanding of the relationship between the CSFH and LGRBR would be incredibly valuable. However one obstacle that plagues the CSFH-LGRB rate connection is whether or not LGRBs are biased tracers of star formation. The question of biasing was first introduced because core-collapse models could not generate a LGRB without the progenitor system having low-metallicity \citep[$\approx0.3\,\mathrm{Z_{\sun}}$; e.g.,][]{Hirschi05, Yoon05, Woosley06}. Also, it was noticed that the LGRBR was flatter at higher redshifts than the CSFH and that LGRBs were typically found in low-mass, low-metallicity galaxies \citep[see e.g.,][]{LeFloch03, Savaglio09, Sven10, Mannucci11}. The difference in the LGRBR and CSFH, suggested in previous papers, can be understood if cosmic metallicity thresholds, evolving LGRB luminosity distributions \citep[e.g., the progenitor evolves with redshift;][]{Virgili11}, evolving stellar initial mass functions \citep{Wang11} or sample selection effects \citep[see e.g.,][]{Coward08, Lu11} are included.

A significant hindrance of the analysis of the CSFH-LGRBR connection is the number of LGRBs with no redshift measurement, i.e., redshift incompleteness. For a redshift to be measured, high precision localisations ($\sim$0.5\arcsec) are required for follow-up spectroscopy. Since the advent of the {\it Swift} satellite \citep{Swift04}, a survey telescope equipped with a GRB alert telescope (BAT), an X-ray telescope (XRT) and a ultra-violet/optical telescope (UVOT), there have been over 600 GRB detections in the past 7 years. The combination of large sky coverage of the BAT and the (near-) instantaneous follow-up with the XRT and UVOT, high precision localisations have been achievable, which were not as common in the preceding era. Accompanying this, the number of robotic ground-based follow-up telescopes has increased, allowing GRBs to be seen minutes and even seconds \citep{Rykoff09} after they trigger the BAT. However, despite such efforts it has only been possible to obtain redshifts for $\sim30\%$ of GRBs in comparison to the high success (84\%) of X-ray detections\footnote{Taken from the {\it Swift} GRB tables; \url{http://heasarc.nasa.gov/docs/swift/archive/grb_table/stats/}}.

Such a low redshift completeness is the result of: large uncertainties in the GRB localisations, weather, the GRB does not fit the redshift follow-up program criteria and GRB sky location, to name but a few. This biasing can become so complex, that it is not always so simple to remove from the sample considered, but none the less has been tried before \citep[e.g.,][]{Coward08}. Consequently, many groups have been trying to improve the completeness levels of their statistics. One such instrument has been set to this task, the gamma-ray burst optical near-infrared (NIR) detector \citep[GROND;][]{GROND}. This multi-channel imager mounted at the 2.2m MPG/ESO telescope at La Silla (Chile) has operated for the past 4 years as an automated GRB afterglow follow-up instrument. The robotic nature of GROND has allowed it to significantly increase the afterglow detection rate ($\sim98\%$) for all LGRBs observed within 4 hours of the trigger. Due to the reduced attenuation from gas and dust at observer frame NIR wavelengths, GROND's multi-band capabilities in combination with its rapid-response allows photometric redshifts to be determined when spectroscopic observations were not possible or fruitless (i.e., redshift desert, low signal-to-noise). This has facilitated, for the first time, highly complete GRB redshift samples.

Only recently have highly complete samples been used to show that GRBs that exhibit no afterglow are primarily the result of these GRBs originating in a galaxy of high extinction \citep{Greiner11}. These {\it dark bursts} \citep[see also][]{Fynbo09, Cenko09, Perley09} are usually not included in GRB host follow-up programs due to poor localisations. New evidence has shown that GRBs with heavily dust-extinguished afterglows exist in systematically more massive galaxies \citep{Kruehler11}, thus originally biasing our opinion on host galaxies. Furthermore, it has been seen before that not all GRB host galaxies fit the picture of low-mass, low-metallicity \citep[see e.g.,][]{Levesque10, Hashimoto10, Savaglio11} and that some can occur in extremely dust extinguished galaxies \citep[][Rossi et al. 11]{Levan06, Berger07, Hashimoto10, Hunt11}, which are also usually associated with large metallicities. 

In this work we investigate, simultaneously, the effect of mass ranges and metallicity cuts placed on the CSFH and its effects on the CSFH-LGRBR connection with the recently available and highly redshift-complete sample of \citet{Greiner11}. The paper is outlined as follows. The CSFH and LGRBR models are explained in Sect.~\ref{sec:CosmicStarFormationAndGRBRate}. The description of individual LGRBs and their properties are discussed in Sect.~\ref{sec:GrondAndDataSamples}. The implementation of the model with the data is presented in Sect.~\ref{sec:Methodology}. The LGRB host property results and differences to other studies are discussed in Sect.~\ref{sec:Results}. In Sect.~\ref{sec:Discussion} we discuss the caveats of our approach and make predictions on the LGRB distribution and CSFH at high redshifts, and summarise our conclusions in Sect.~\ref{sec:Conclusion}. A $\Lambda$CDM cosmology has been assumed throughout this paper ($\Omega_{\Lambda}=0.7$, $\Omega_{\mathrm{M}}=0.3$, $H_{0}=73.0\,\mathrm{km\,s^{-1}\,Mpc^{-1}}$).
\section{CSFH \& LGRB Rate Models}
\label{sec:CosmicStarFormationAndGRBRate}

The aim of this work is to compare LGRB number densities generated by models, that vary with regards to their host galaxy characteristics, to an observed LGRB number density. To do this, the CSFH is converted to a LGRBR utilising a conversion factor, $\eta_{\mathrm{grb}}$ (i.e., from the stars in a given galaxy how many will produce a LGRB that is then observed by a given instrument). The number density is then calculated by numerical integration. The whole process is outlined in the next section and is divided in the following way:
\begin{enumerate}[A.]
 \item { \noindent{\bf CSFH Model}

	\begin{enumerate}[1.]

	\item The CSFH is formulated from empirically constrained models of galaxy star formation rates~(Sect.~\ref{subsec:SFR}) and galaxy mass functions~(Sect.~\ref{subsec:GalaxyMassFunction}).

	\item Restrictions are implemented in the CSFH model~(Sect.~\ref{subsec:CosmicStarFormationRate}) on such things as: galaxy mass ranges~(Sect.~\ref{subsec:GalaxyMassFunction}), galaxy metallicities~(Sect.~\ref{subsec:MassMetallicityRelation}) and red-dead galaxies~(Sect.~\ref{subsec:GalaxyDownsizing}).

	\end{enumerate}
	}

 \item { \noindent{\bf LGRBR Model}
	\begin{enumerate}[1.]

	\item An initial mass function for stars is chosen~(Sect.~\ref{subsec:InitialMassFunction}), which will be used for $\eta_{\mathrm{grb}}$.

	\item A LGRB luminosity function is chosen to model the samples that will be investigated and implemented in $\eta_{\mathrm{grb}}$~(Sect.~\ref{subsec:GRBLuminosityFunction}).

	\item The LGRB number density is calculated using numerical integration and $\eta_{\mathrm{grb}}$~(Sect.~\ref{subsec:GammaRayBurstRate}).

	\end{enumerate}
	}

\end{enumerate}

\subsection{Star Formation Rate}
\label{subsec:SFR}

During a galaxy's evolution, its gas supply will go through two distinct phases named the {\it source} and {\it sink}. The source phase is the inflow (accretion) of cold gas due to the halo's potential well, and the sink phase is the consumption/loss of gas by production of stars and outflows. These two processes occur constantly during the evolution of a galaxy with different weighting, e.g., at redshift 2 it is still possible to have cold accretion on to massive galaxies \citep[see e.g.,][]{Khochfar09}. However, it has been shown via simulations that galaxies will reach a steady state between these two phases \citep{Bouche} and that the SFR can be considered to be solely dependent on the inflow of cold gas. It is this process that is believed to result in the SFR ``sequence'', i.e., the relationship between the star formation rate and stellar mass, $\mathrm{M_{*}}$, of a galaxy. This sequence is seen to obey the following relationship \citep{Bouche}:

\begin{equation}
SFR\left(M_{*}, z\right) = 150\left(M_{*}/10^{11}M_{\sun}\right)^{0.8}\left(\frac{1+z}{3.2}\right)^{2.7} \mathrm{M_{\sun}yr^{-1}}.
\label{eqn:sfr}
\end{equation}

\noindent The main drawbacks of this formulation are that it is only fit up to redshifts of $z=2$ \citep[for a model that can account for high redshift observations of the SFR stellar mass relation, see][]{Kochfar11} and that it is built from average stellar masses. This results in a spread of $\sim0.3\,\mathrm{dex}$. The final model is compared to data in Sect.~\ref{subsec:CosmicStarFormationRate}.


\subsection{Galaxy Mass Function}
\label{subsec:GalaxyMassFunction}
The galaxy mass distribution function (GMF) is commonly described by a Schechter function \citep{Schechter76}. However, as galaxies evolve they accrete more gas and create more stars, thus modifying the number density of galaxies at a given mass for a specific redshift \citep[e.g.,][]{Khochfar09b}. Therefore, a redshift evolving GMF, as measured from the GOODS-MUSIC field \citep{Fontana}, is used in this paper \citep[similarly to][]{Young07, Belczynski10}:

\begin{equation}
\phi\left(\mathcal{M}, z\right) = \phi^{*}\left(z\right)\ln\left(10\right)[10^{\mathcal{M}-M^{*}\left(z\right)}]^{1+\alpha^{*}\left(z\right)}e^{-10^{\mathcal{M}-M^{*}\left(z\right)}},
\label{eqn:gmf}
\end{equation}
where $\mathcal{M}=\log_{10}\left(M_{*}/M_{\sun}\right)$, $M_{*}$ is the stellar mass of the galaxy and the parametric functions obey:
\[
\phi^{*}\left(z\right) = \phi^{*}_{0}\left(1+z\right)^{\phi_{1}^{*}}
\]
\[
\alpha^{*}\left(z\right) = \alpha_{0}^{*}+\alpha_{1}^{*}z
\]
\[
M^{*}\left(z\right) = M_{0}^{*} + M_{1}^{*} z + M_{2}^{*} z^{2}
\]

\noindent The parameter values are given in Table~\ref{table:gmf_param}. Given that the above GMF is modelled on flux limited surveys, we note that methods have been used to calculate missing galaxies by mass-to-light ratios. Secondly, the empirical fits are also only applicable for redshifts, $z<4$. See \citet{Fontana} for full details on the galaxy selection and GMF fitting.

\begin{table}[h!]
 \caption{Galaxy mass function parameters taken from \citet{Fontana}.}

  \centering
    \begin{tabular}{ l l l l }
    \hline
      $M_{0}^{*}$ & $M_{1}^{*}$ & $M_{2}^{*}$ & $\alpha^{*}_{0}$ \\
    \hline
      & & & \\
    \hline
      11.16 & 0.17$\pm$0.05 & -0.07$\pm$ 0.01 & -1.18 \\ 
    \hline
    \hline
      & & & \\
      $\alpha^{*}_{1}$ & $\phi^{*}_{0}$ & $\phi^{*}_{1}$ & \\
    \hline
      & & & \\
    \hline
      -0.082$\pm$0.033 & 0.0035 & -2.20$\pm$0.18 & \\
    \hline\hline
   \end{tabular}
  \label{table:gmf_param}

\end{table}

\subsection{Mass-Metallicity Relation}
\label{subsec:MassMetallicityRelation}
As galaxies evolve and stars form, they pollute the galaxy with metals through stellar winds, supernovae and other forms of feedback and therefore can mediate the inflow and collapse of gas. A mass-redshift dependent metallicity empirically fit relation, $\epsilon$, is assumed throughout this paper, taken from \citet{Savaglio05}, of the form\footnote{The conversions $\epsilon\left[O/H\right]+12=\epsilon\left[Z/\mathrm{Z_{\sun}}\right]+8.69$, \citep{Savaglio06} and $t_{H}=\frac{\left(1+z\right)^{-1.5}}{{1.4\sqrt{\Omega_{\Lambda}}H_{0}}}$ are used throughout this paper.}:

\begin{eqnarray}
  \mathrm{\epsilon\left[O/H\right]} = \epsilon\left(M_{*},t_{H}\right) = -7.5903 + 2.5315 \log M_{*} \nonumber \\ 
  - 0.09649 \log^{2} M_{*} + 5.1733 \log t_{H} \nonumber \\
  - 0.3944 \log^{2} t_{H} - 0.0403 \log t_{H} \log M_{*}
\label{eqn:metallicity}
\end{eqnarray}

\noindent where $\mathrm{M_{*}}$ is the stellar mass and $t_{H}$ the Hubble time measured in $10^{9}$ years. The relation is taken to be true over the redshift range we investigate (up to $z\approx12$), but has been noted that it is difficult to reconcile for $z>3.5$ \citep{Maiolino08}


\subsection{Galaxy Downsizing}
\label{subsec:GalaxyDownsizing}

Massive galaxies in the redshift regime of $z\sim1-3$ have been seen to have little (passively evolving) or no (dead) star formation \citep[][]{Cimatti04, Labbe05, Daddi05, Kong06}. Observationally, more massive galaxies are seen to cease star formation first and this has been coined downsizing. The mechanism to explain this process is believed to be caused by the quenching of star formation by active galactic nuclei \citep[AGN; e.g.,][]{Croton06}, galaxy mergers \citep[e.g.,][]{Springel05}, temperature thresholds \citep[e.g.,][]{Dekel06} or gravitational heating \citep[e.g.,][]{Khochfar08}, but the relative importance is still unclear. It is possible to experimentally constrain the quenching stellar mass, $M_{\mathrm{Q}}$, above which no star formation is occurring. This is also seen to evolve strongly with redshift in the following manner \citep{Bundy06}:

\begin{equation}
 M_{\mathrm{Q}}\left(z\right) = M_{\mathrm{Q}}^{0}\left(1+z\right)^{3.5},
\label{eqn:quenchingmass}
\end{equation}

\noindent where $M_{\mathrm{Q}}^{0}$ is the local ($z\sim0$) quenching mass. A value of $\log_{10} \left(M_{\mathrm{Q}}^{0}/\mathrm{M_{\sun}}\right)=10.9\pm0.1$ \citep{Gallego11} is assumed throughout.

\subsection{CSFH}
\label{subsec:CosmicStarFormationRate}

Given the previous models, it is possible to model the CSFH. This is simply the integration of the individual galaxy SFRs over the GMF. As our aim was to also investigate varying characteristics of galaxies that contribute to the CSFH, we include the following free parameters: metallicity upper limit, $\epsilon_{\mathrm{L}}$, and stellar mass upper and lower limits, $M_{1},M_{2}$. This results in a CSFH of the form,

\begin{eqnarray}
   \dot{\rho} & = & \int_{M_{1}}^{M_{2}}\zeta\left(z\right)\gamma\left(M_{*},z,\epsilon_{\mathrm{L}}\right)SFR\left(M_{*}, z\right) \phi\left(M_{*},z\right)\, \mathrm{d}M_{*} \nonumber\\ 
   & = & \dot{\rho}\left(z,\epsilon_{\mathrm{L}},M_{1},M_{2}\right) 
 \label{eqn:csfr}
\end{eqnarray}

\noindent where the metallicity and downsizing constrains $\gamma$ and $\zeta$ respectively, are defined by the following relations:

\begin{equation}
   \gamma\left(M_{*},z,\epsilon_{\mathrm{L}}\right) = \left\{
     \begin{array}{l l}
 	1 & \mathrm{if}\,\epsilon\left(M_{*},z\right)<\epsilon_{\mathrm{L}} \\
	0 & \mathrm{if}\,\epsilon\left(M_{*},z\right)\ge\epsilon_{\mathrm{L}} \\
     \end{array} 
   \right.
 \label{eqn:gammafnc}
\end{equation}

\begin{equation}
  \mathrm{\zeta(z)} = \left\{
      \begin{array}{l l}
       1 & \quad \mathrm{if} \,M_{\mathrm{Q}}\left(z\right)> M_{*} \\
       0 & \quad \mathrm{if} \,M_{\mathrm{Q}}\left(z\right)\le M_{*} \\
      \end{array}
  \right.
 \label{eqn:zetafnc}
\end{equation}

\begin{figure}
\resizebox{\hsize}{!}{\includegraphics[width=7.5cm]{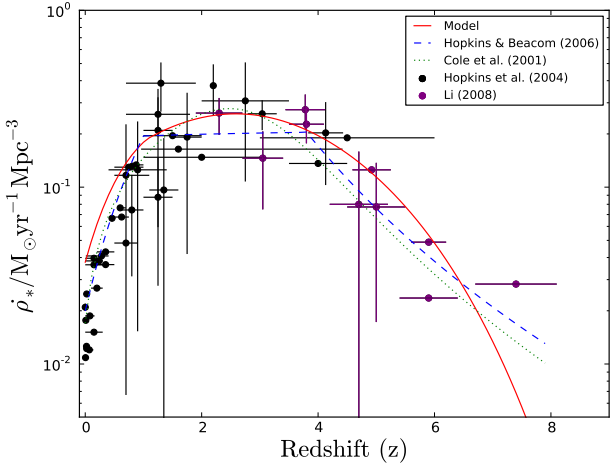}}
\caption{The CSFH given by the model in Eqn.~\ref{eqn:csfr} using no metallicity limit (i.e., no $\epsilon_{\mathrm{L}}$) and $(\mathcal{M}_{1}\mathcal{M}_{2})=(7,12)$ (bold-red line), and two empirical fits (green-dotted and blue-dashed lines). The black circles denote data from \citet{Hopkins04} and purple from \citet{Li08}.}
\label{fig:csfr_mod_emp}
\end{figure}

The form of the above two step-functions (Eqns.~\ref{eqn:gammafnc} and \ref{eqn:zetafnc}) could be different. For example, \citet[][]{Wolf07} implemented a metallicity cut-off of not only step-functions, but power laws and broken power laws. Their results show that they cannot currently discriminate between the form the metallicity cut can take. Secondly, it has been seen observationally that the quenching mass is not a single cut-off but a smooth function of mass \citep[e.g.,][]{Kong06}, however, the knowledge of this trend is also not well known. Therefore, for simplicity both are chosen to be implemented as step-functions.

The numerically integrated model, presented in Fig. \ref{fig:csfr_mod_emp}, is compared to data from \citet{Hopkins04} and \citet{Li08}, and to two empirical fits from \citet{HopkinsBeacom06} and \citet{Cole01}. The behaviour of the three curves is similar, peaking between redshift 2 and 3 and becoming negligible above redshift 7.  A goodness of fit of our model to the data yields a $\chi^{2}/\df=$36/50. There is a small overproduction of stars at low redshift in comparison to the data, which can be lowered by reducing the quenching mass, $M_{Q}^{0}$. However, there is a range of possible quenching masses that could be used and still give a good fit to the experimental data. As a result, it was opted to use the local quenching mass given by \citet{Gallego11}. It is noted that this parameter could also have been left free in our analysis, however, simulations we ran using no quenching mass gave the same results that are outlined in Sect.~\ref{sec:Results}.

The important difference between the approach used in this paper to others is that the form of the model in Eqn.~\ref{eqn:csfr} allows freedom for the mass and metallicity of the contributing host galaxies. Also, as we are interested in the low-redshift range ($z<3$), we can remove the freedom of the parameters of the individual models incorporated (e.g., Eqn.~\ref{eqn:sfr}.).


\subsection{Initial Mass Function}
\label{subsec:InitialMassFunction}
Previously, it was mentioned that different types of initial mass functions (IMFs) have been used to explain the current LGRBR-CSFH connection. Currently there is no consensus on the correct IMF to be used: Salpeter \citep{Salpeter55}, Scalo \citep{Scalo86}, Kroupa \citep{Kroupa01} or Chabrier \citep{Chabrier03}, and is still a lively debated issue \citep[see][]{Bastian10}. A commonly chosen IMF is the Salpeter IMF and is used throughout this work:

\begin{equation}
\psi\left(m\right) = m^{-\alpha},
\label{eqn:imf}
\end{equation}
\noindent where $m$ is the star mass and $\alpha=2.35$. We note that the collapsar model requires the formation of a black hole and this only applies to the high end mass of stars ($>30 \mathrm{M_{\sun}}$). The main difference between possible IMFs is at the low-mass end, and therefore any IMF chosen would have been viable as long as it was redshift independent. We do not consider an evolving IMF~\citep[top heavy; see e.g.,][]{Dave08, Weidner11}, but similar studies have been done utilising this type \citep[e.g.,][]{Wang11}.


\subsection{GRB Luminosity Function}
\label{subsec:GRBLuminosityFunction}
The LGRB luminosity function (LF) can take on many forms, for example: Schechter functions, broken power-laws, log-normal functions and normal functions (Gaussian function). Throughout this paper a normal function has been used \citep[see e.g.,][]{Bastian10, Belczynski10} of the form:

\begin{equation}
\label{eqn:GRBLF}
\phi\left(L\right) = n_{*} e^{\left(L-L_{\mathrm{c}}\right) / 2\sigma_{L}^{2}},
\end{equation}

\noindent where $L=\log\left(L_{iso} / \mathrm{erg\,s^{-1}}\right)$. One of the fits to our data can be seen in Fig. \ref{fig:GRDB_LuminosityRedshiftSample:b}, and the best-fits for all the samples considered (see Sect.~\ref{sec:GrondAndDataSamples}) can be found in Table~\ref{tab:grblf}. The total fraction of LGRBs seen by a specific instrument, then takes the form:

\begin{equation}
 \label{eqn:NGRBLF}
 f\left(z\right) = 
\frac{
\int_{L_{\mathrm{limit}\left(z\right)}}^{\infty} \phi\left(L\right)\mathrm{d}L
}
{
\int_{-\infty}^{\infty} \phi\left(L\right)\mathrm{d}L
}.
\end{equation}

\noindent The luminosity limit, $L_{\mathrm{limit}\left(z\right)}$, of the sample can be calculated using the luminosity distance, $D_{L}$, of the form $L_{\mathrm{limit}} = 4\pi D_{L}^2F_{\mathrm{limit}}$. By taking $F_{\mathrm{limit}}$ to be the lowest luminosity of the sample (see Sect.~\ref{sec:GrondAndDataSamples}), results in the following flux limit: $F_{\mathrm{limit}}=1.08\cdot10^{-8}\,\mathrm{erg\,s^{-1}\,cm^{-2}}$. This flux limit is essentially the limiting flux of the BAT on board the {\it Swift} satellite. 

It is also possible that the LGRB LF is a function of redshift. A redshift dependence would imply that there is a change in the LGRB explosion mechanism throughout redshift or that the progenitor mass distribution, or mode (single, binary), is changing. This has been investigated previously by \citet{Salvaterra07} and \citet{Campisi10} who both come to the conclusion that a non-evolving LF is possible, but requires a metallicity cut to reproduce the observed LGRB distribution (for metallicity cuts see Sect.~\ref{sec:Methodology}). On the contrary to this, \citet{Butler10} rule out an evolving luminosity function to the $5\sigma$ level using {\it Swift} data. To increase the simplicity of the modelling, we assumed that the mechanism was redshift independent and that the progenitor was of a constant type. For a discussion on implications of these assumptions, see Sect.~\ref{subsec:EmpiricalLawsAssumptionsDegeneracy}.

\begin{table}
\caption{GRB Luminosity Function best-fit parameters.}
  \centering
  \begin{tabular}{ l  l  l  l  l }
	  \hline
		  Sample & $\log_{10}\left(L_{\mathrm{c}}/\mathrm{erg\,s^{-1}}\right)$  & $\sigma_{L}$ & $n_{*}$ & $\chi^{2}/\df$ \\ 
	  \hline
		  & & & & \\
	  \hline
		  S & 51.33 & 0.97 & 10.89 & 0.63/4 \\
		  P & 51.74 & 0.93 & 13.19 & 5.56/5 \\
		  UL1 & 51.06 & 0.87 & 23.48 & 0.52/3 \\
		  UL2 & 51.11 & 0.92 & 22.95 & 0.39/3 \\
	  \hline\hline

   \end{tabular}
\label{tab:grblf}
\end{table}

\subsection{Long Gamma-Ray Burst Rate}
\label{subsec:GammaRayBurstRate}
Given the CSFH model formulated in the previous section, we now consider which population of galaxies contribute to the observed LGRBR. To do this, six parameters have to be considered; (i) the co-moving volume; (ii) the instruments limiting flux, that affects how many LGRBs of a given luminosity can actually be detected; (iii) a LGRB probability, that transforms the number of stars being formed to the number of LGRBs being produced; (iv) proportionality relation of the CSFH-LGRB rate, to include any further dependencies unaccounted for in our model; (v) the fraction of the sky observable by the instrument; (vi) and the length of time the instrument has been running. The total detectable LGRB rate for a given instrument in a redshift bin $\left(z_{1}, z_{2}\right)$ is then simply:

\begin{eqnarray}
N\left(z_{1},z_{2}\right) = \eta_{\mathrm{grb}}\int_{z_{1}}^{z_{2}} \frac{f\left(z\right)\dot{\rho}\left(z,\epsilon_{L},M_{1},M_{2}\right)\left(1+z\right)^{\delta} \frac{\mathrm{d}V}{\mathrm{d}z}}{\left(1+z\right)}\,\mathrm{d}z,
\label{eqn:grb_rate}
\end{eqnarray}
\noindent where $\frac{\mathrm{d}V}{\mathrm{d}z}=4\pi D_{\mathrm{com}}^{2}\left(z\right)\frac{\mathrm{d}D_{\mathrm{com}}\left(z\right)}{\mathrm{d}z}$, $D_{\mathrm{com}}$ is the co-moving distance, $\eta_{\mathrm{grb}}$ is the probability of stars resulting in a LGRB and then detected by an instrument,  and $\delta$ is the power of proportionality for the CSFH to LGRBR \citep[see e.g.,][]{Bromm06, LangerNorman06, Daigne06, Young07, Wolf07, Salvaterra07, Kistler09, Campisi10, Qin10, Butler10, Wanderman10, Belczynski10, Virgili11, deSouza11, Wang11, Ishida11}. The LGRB probability can be parameterised as \citep[explained similarly in][]{Bromm06}:

\begin{equation}
	\eta_{\mathrm{grb}} = \Delta T\,\Delta \Omega\,\eta_{\mathrm{coll}}\,\eta_{\mathrm{BH}}\,\eta_{\mathrm{time}}\,\eta_{\mathrm{X-ray}}\,\eta_{\mathrm{redshift}}\,\eta_{\mathrm{other}},
	\label{eqn:grbprob}
\end{equation}

\noindent where $\Delta T$ is the length of observations, $\Delta\Omega$ is the solid angle of the observable sky and the remaining parameters are discussed in the following paragraphs.

As mentioned, LGRBs are likely associated with black holes produced by the collapse of stars above a specific progenitor mass, $M_{\mathrm{BH}}$, and is quantified in the following way:
\begin{equation}
 \eta_{\mathrm{BH}} = \frac{\int_{M_{\mathrm{BH}}}^{M_{\mathrm{max}}} \psi\left(m\right)\,\mathrm{d}m}{\int_{M_\mathrm{min}}^{M_{\mathrm{max}}} m\psi(m)\,\mathrm{d}m},
 \label{eqn:bheff}
\end{equation}
where $M_{\mathrm{min}}$ and $M_{\mathrm{max}}$ are the minimum and maximum star masses considered in our model respectively. Secondly, GRBs are believed to be jets of collimated matter, as a result of breaks observed in afterglow light curves. This means that only a small fraction of bursts are visible to the observer, quantified by $\eta_{\mathrm{coll}}$. 

The last four probabilities, $\eta_{\mathrm{time}}$, $\eta_{\mathrm{X-ray}}$, $\eta_{\mathrm{redshift}}$ and $\eta_{\mathrm{unknown}}$ are a result of our sample selection criteria, outlined in the next section. They are the probability of detecting an optical-NIR afterglow less than 4 hours after a GRB trigger, the probability of detecting an X-ray afterglow, the probability of detecting a redshift, and any unknown probabilities respectively.

Finally, the parameter $\delta$ includes further dependencies between the CSFH and LGRBR not accounted for by our model. This parameter is the same utilised in \citet{Kistler09}, where it is used to incorporate effects that are not known, i.e., a black-box approach. We employ the same idea, so that a non-zero $\delta$ would imply we are missing an effect in our CSFH modelling. One example is an evolving LGRB LF or an evolving stellar IMF. It is now possible to formulate a LGRB number density distribution for a redshift bin through the use of a modelled CSFH, which can then be compared to an experimental data set. We would like to note that the empirically calibrated relations used in the model are only verified for redshifts $z<4$, and could differ at higher redshifts. To investigate the implications of this on our final results, we manually modify the CSFH at redshifts $z>3$ in Sect.~\ref{Results:High-zCSFR}.
\section{Data Samples}
\label{sec:GrondAndDataSamples}

\subsection{Gamma-ray Burst Sample}
\label{subsec:GrondSamplesDarkBurstsSamples}

Our LGRB sample is taken from \citet[][]{Greiner11}, which unlike previous studies, is highly complete in terms of optical/NIR afterglow detection rates and measured redshifts. \citet[][]{Greiner11} chose the sample by selecting GRBs that have been detected by GROND within 4 hours after the {\it Swift} BAT trigger and that exhibited an X-ray afterglow. This selection results in a sample of 43 GRBs: 39 LGRBs and 4 short GRBs \citep[believed to be associated with the merger of 2 neutron stars or a neutron star-black hole system;][]{Belczynski06}. The 39 LGRB sample contains 31 spectroscopic redshifts, 6 photometric redshift measurements (3 of which are upper limits) and 2 with no optical/NIR afterglow detections and thus no redshift measurements.  For more details on the burst sample and individual bursts, see \citet[][]{Greiner11}.

\begin{figure}
  \includegraphics[width=9cm]{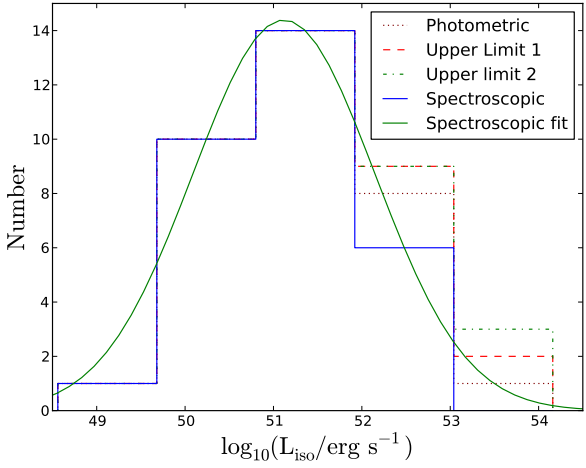}
  \caption{Number density of the luminosities for all of the subsamples considered. The bold line depicts an example best-fit normal-Gaussian to the spectroscopic sample. Bin sizes are chosen for presentation of the data.}
  \label{fig:GRDB_LuminosityRedshiftSample:b}
\end{figure}

\begin{figure}
    \includegraphics[width=9cm]{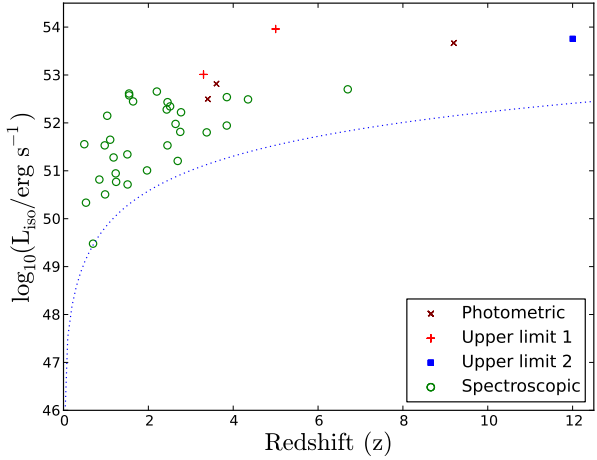}
    \caption{Luminosity-redshift space of all the subsamples considered. The dashed line depicts the flux limit, found by fitting the spectroscopic distribution, $F_{\mathrm{limit}}=1.08\cdot10^{-8}\mathrm{\,erg\,s^{-1}\,cm^{-2}}$ (see Sect.~\ref{subsec:GRBLuminosityFunction}).}
    \label{fig:GRDB_LuminosityRedshiftSample:a}
\end{figure}

\begin{figure}
  \includegraphics[width=9cm]{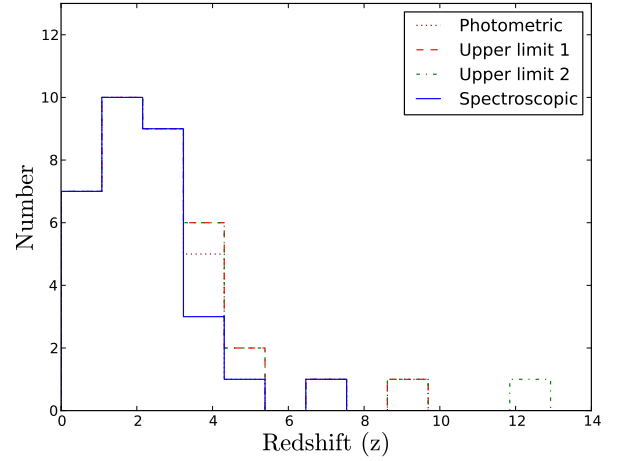}
  \caption{Number density of the redshifts for all of the subsamples considered. Bin sizes are chosen for presentation of the data.}
  \label{fig:GRDB_LuminosityRedshiftSample:c}
\end{figure}

As mentioned previously, it is only LGRBs that are believed to trace the death of massive stars, and so the short-GRBs were not considered in our analysis. The LGRB sample was then subdivided into; spectroscopic (S):  bursts with spectroscopic redshift (31/39); photometric (P): spectroscopic sample including photometric redshifts (34/39); upper limit 1 (UL1): photometric sample including upper limit photometric redshifts (36/39); and upper limit 2 (UL2): upper limit sample including a possible redshift measurement (37/39). This subdivision was introduced for two reasons. Firstly, to see if the results changed when having a single method (spectroscopic) of redshift identification compared to multiple types of redshift identification (spectroscopic and photometric). Secondly, the upper limits were included to get as close to a 100\% redshift-detected-sample as possible.
All the samples cover a range of $z=0$ to $z=6.7$, and the P, UL1 and UL2 samples cover $z=0$ to $z=9.2$. Any bias in the redshift distribution, i.e., possible deficits in the high redshift regime and effects of selecting bursts followed by GROND $<4$ hours after the trigger, has been investigated by \citet{Greiner11}. This was carried out by comparing the sample to one of twice the size \citep{Fynbo09}, with a completeness at the time of 50\%, utilising a Kolmogorov-Smirnov test. This showed that the difference at high-z was statistically insignificant and at the $1\sigma$ level, both samples could be drawn from the same underlying distribution. For information on individual bursts see Table~\ref{tab:Samples:GrondDarkBurstSamples} and for the property-distributions see Figs.~\ref{fig:GRDB_LuminosityRedshiftSample:b},~\ref{fig:GRDB_LuminosityRedshiftSample:a} and \ref{fig:GRDB_LuminosityRedshiftSample:c}.

\subsection{LGRB Properties}
\label{subsec:DataSamplesProperties}

For each burst the measurement of a redshift and luminosity is required for the modelling of the CSFH-LGRBR, which will be discussed in Sect.~\ref{sec:CosmicStarFormationAndGRBRate}. Redshifts for all of the samples were obtained from the table compiled by \citet{Greiner11} and the luminosities were calculated from the isotropic energy ($E_{\mathrm{iso}}$) and the duration of the burst over which 90\% of the flux is released ($T_{90}$), using the standard relation of $L_{\mathrm{iso}} = E_{\mathrm{iso}}\left(1+z\right)/T_{90}$ \citep{Bloom01}. The $E_{\mathrm{iso}}$ and $T_{90}$ for each burst was obtained from fits to the prompt emission taken from the extension of \citet[][from here NB11]{Butler07}\footnote{\url{http://astro.berkeley.edu/~nat/swift/}}. For bursts with photometric or upper limit redshifts an $E_{\mathrm{iso}}$ was calculated via the same procedure described in \citet{Bloom01}. One of the following energy spectra, $\phi\left(\mathrm{E}\right)$, was used based on the best-fit to the LGRB's BAT spectra:

\begin{equation}
  \phi\left(E\right) = \left\{
    \begin{array}{l l}
     K_{A}\,\left(\frac{E}{A}\right)^{\alpha} & \quad \textrm{Power\,Law\,(PL)}\\
      & \\
     K_{A}\,\left(\frac{E}{A}\right)^{\alpha}e^{\frac{-E(2+\alpha)}{E_{\mathrm{peak}}}} & \quad \textrm{Cut-off\,Power\,Law\,(CPL)}\\
    \end{array} \right.,
\end{equation}\noindent~\citep[see e.g.][]{Sakamoto08} where $K_{A}$ is the normalisation at $E=A\,\mbox{keV}$ in units of $\mathrm{photons\,keV^{-1}\,cm^{-2}s^{-1}}$, $E_{\mathrm{peak}}$ is the peak energy in the cut-off power-law spectrum, i.e., in $E^{2}\phi\left(E\right)$ space, and $\alpha$ is the spectral slope. Utilising the best-fit model, the K-corrected $E_{\mathrm{iso}}$ is calculated via\footnote{For BAT it is assumed $E_{1}=15\,\mathrm{keV}$ and $E_{2}=150\,\mathrm{keV}$, but we rescale similar to NB11 to the range of, $E_{1}=1\,\mathrm{keV}$ to $E_{2}=10^{4}\,\mathrm{keV}$}:

\begin{equation}
E_{\mathrm{iso}} = \Delta t\frac{4\pi D_{L}^{2}}{1+z} \int^{E_{2}/\left(1+z\right)}_{E_{1}/\left(1+z\right)} E\phi\left(E\right)\,\mathrm{d}E,
\end{equation}
\noindent where $D_{L}$ is the luminosity distance, $z$ the redshift, and $\Delta t$ the time over which the spectra is fit (a time-integrated spectra).

\begin{table*}
  \caption{The properties of the GROND Dark Burst samples.}
  \label{tab:Samples:GrondDarkBurstSamples}
  \centering

  \begin{tabular}{ l l l l l l l l l l }
    \hline
      Burst & $z$\tablefootmark{a} & $\log_{10}\left(E_{\mathrm{iso}}/\mathrm{erg}\right)$ & $T_{90}$ & \multicolumn{5}{ c }{Prompt Model~\tablefootmark{b}} & Sample~\tablefootmark{c} \\
      \cline{5-9}
      & & & & Name & $\alpha$ & $E_{\mathrm{peak}}$ & $K_{A}$ & $\Delta t$ & \\
    \hline
      & & & (s) & & & (keV) & $\mathrm{ph\,keV^{-1}\,cm^{-2}s^{-1}}$ & (s) & \\
    \hline
      GRB100316B      & 1.180 (1) & 51.57        & 4.3 & ... & ... & ... & ... & ... &  S, P, UL1, UL2 \\
      GRB091127       & 0.490 (2) & 52.36        & 9.6 & ... & ... & ... & ... & ... &  S, P, UL1, UL2 \\
      GRB091029       & 2.752 (3) & 52.84        & 40.0 & ... & ... & ... & ... & ... &  S, P, UL1, UL2 \\
      GRB091018       & 0.971 (4) & 51.88        & 4.4 & ... & ... & ... & ... & ... &  S, P, UL1, UL2 \\
      GRB090926B      & 1.24 (5) & 52.52        & 126.4 & ... & ... & ... & ... & ... &  S, P, UL1, UL2 \\
      GRB090814A      & 0.696 (6) & 51.30        & 113.2 & ... & ... & ... & ... & ... &  S, P, UL1, UL2 \\
      GRB090812       & 2.452 (7) & 53.89        & 99.8 & ... & ... & ... & ... & ... &  S, P, UL1, UL2 \\
      GRB090519       & 3.85 (8) & 53.76        & 81.8 & ... & ... & ... & ... & ... &  S, P, UL1, UL2 \\
      GRB090313       & 3.375 (9) & 53.11        & 90.2 & ... & ... & ... & ... & ... &  S, P, UL1, UL2 \\
      GRB090102       & 1.547 (10) & 53.65        & 30.7 & ... & ... & ... & ... & ... &  S, P, UL1, UL2 \\
      GRB081222       & 2.77 (11) & 53.17        & 33.5 & ... & ... & ... & ... & ... &  S, P, UL1, UL2 \\
      GRB081121       & 2.512 (12) & 53.08        & 19.4 & ... & ... & ... & ... & ... &  S, P, UL1, UL2 \\
      GRB081029       & 3.848 (13) & 53.48        & 169.1 & ... & ... & ... & ... & ... &  S, P, UL1, UL2 \\
      GRB081008       & 1.968 (14) & 52.83        & 199.3 & ... & ... & ... & ... & ... &  S, P, UL1, UL2 \\
      GRB081007       & 0.529 (15) & 50.89        & 5.6 & ... & ... & ... & ... & ... &  S, P, UL1, UL2 \\
      GRB080913       & 6.7 (16) & 52.72        & 8.2 & ... & ... & ... & ... & ... &  S, P, UL1, UL2 \\
      GRB080805       & 1.505 (17, 18) & 52.99        & 111.8 & ... & ... & ... & ... & ... &  S, P, UL1, UL2 \\
      GRB080804       & 2.20 (19, 18) & 53.94        & 61.7 & ... & ... & ... & ... & ... &  S, P, UL1, UL2 \\
      GRB080710       & 0.845 (20, 18) & 52.69        & 139.1 & ... & ... & ... & ... & ... &  S, P, UL1, UL2 \\
      GRB080707       & 1.23 (21, 18) & 52.07        & 30.3 & ... & ... & ... & ... & ... &  S, P, UL1, UL2 \\
      GRB080605       & 1.640 (22, 18) & 53.31        & 19.6 & ... & ... & ... & ... & ... &  S, P, UL1, UL2 \\
      GRB080520       & 1.545 (23, 18) & 52.68        & 3.0 & ... & ... & ... & ... & ... &  S, P, UL1, UL2 \\
      GRB080413B      & 1.1 (24, 18) & 52.17        & 7.0 & ... & ... & ... & ... & ... &  S, P, UL1, UL2 \\
      GRB080413A      & 2.433 (25, 18) & 53.41        & 46.7 & ... & ... & ... & ... & ... &  S, P, UL1, UL2 \\
      GRB080411       & 1.03 (26, 18) & 53.60        & 58.3 & ... & ... & ... & ... & ... &  S, P, UL1, UL2 \\
      GRB080330       & 1.51 (27) & 52.13        & 66.1 & ... & ... & ... & ... & ... &  S, P, UL1, UL2 \\
      GRB080210       & 2.641 (28, 18) & 53.06        & 43.9 & ... & ... & ... & ... & ... &  S, P, UL1, UL2 \\
      GRB080129       & 4.349 (29) & 53.42        & 45.6 & ... & ... & ... & ... & ... &  S, P, UL1, UL2 \\
      GRB071031       & 2.692 (30) & 52.91        & 187.2 & ... & ... & ... & ... & ... &  S, P, UL1, UL2 \\
      GRB071010A      & 0.98 (31) & 51.56        & 22.4 & ... & ... & ... & ... & ... &  S, P, UL1, UL2 \\
      GRB070802       & 2.45 (32) & 52.16        & 14.7 & ... & ... & ... & ... & ... &  S, P, UL1, UL2 \\
      \hline
      GRB090429B      & 9.2 (33) & 53.42        & 5.8 & CPL & $-0.69^{+0.91}_{\small-0.76}$ & $46.17^{+6.53}_{-10.72}$ & 0.059 & 7.55 & P, UL1, UL2 \\
      GRB081228       & 3.4 (34) & 52.43        & 3.8 & PL & $-1.99_{-0.35}^{+0.31}$ & ... & 0.028 & 4.44 & P, UL1, UL2 \\
      GRB080516       & 3.6 (35) & 52.98        & 6.8 & PL & $-1.78_{-0.28}^{+0.26}$ & ... & 0.039 & 7.83 & P, UL1, UL2 \\
      \hline
      GRB091221       & $<3.3$\tablefootmark{d} & 54.21        & 69.0 & PL & $	-1.62_{-0.06}^{+0.06}$ & ... & 0.043 & 101.31 & UL1, UL2 \\
      GRB090904B      & $<5.0$\tablefootmark{d} & 54.94        & 58.2 & PL & $-1.58_{-0.08}^{+0.08}$ & ... & 0.105 & 86.40 & UL1, UL2 \\
      \hline
      GRB100205A      & 12.0\tablefootmark{e} & 54.2       & 32.76 & PL & $-1.73_{-0.31}^{+0.29}$ & ... & 0.008 & 41.40 & UL2 \\
    \hline\hline
  \end{tabular}

    \tablefoot{
      \tablefoottext{a}{If not mentioned otherwise, the redshift is taken from \citet{Greiner11}.}
      \tablefoottext{b}{Both the prompt best-fit model and parameters are taken from NB11. They are only listed if the $E_{\mathrm{iso}}$ was calculated manually.}
      \tablefoottext{c}{S: Spectroscopic, P: Photometric, UL1: Upper limit 1, UL2: Upper limit 2.}
      \tablefoottext{d}{These upper limits are treated as being the actual redshift value.}
      \tablefoottext{e}{Value taken from between $11<z<13.5$, \citep{Cucchiara10}.}
      }
  
    \tablebib{
      All references are taken from \citet[][Table 2]{Greiner11}.
      (1) \citet{Vegani10};\,
      (2) \citet{Cucchiara09};\,
      (3) \citet{Chornock09b};\,
      (4) \citet{Chen09};\,
      (5) \citet{Fynbo09b};\, 
      (6) \citet{Jakobsson09};\,
      (7) \citet{deUgartePostigo09b};\,
      (8) \citet{Thoene09};\,
      (9) \citet{Chornock09, deUgartePostigo10b};\,
      (10) \citet{deUgartePostigo09};\,
      (11) \citet{Cucchiara08};\,
      (12) \citet{Berger08b};\,
      (13) \citet{DElia08};\,
      (14) \citet{DAvanzo08};\,
      (15) \citet{Berger08};\,
      (16) \citet{Greiner09};\,
      (17) \citet{Jakobsson08d};\,
      (18) \citep{Fynbo09};\,
      (19) \citet{Thoene08c};\,
      (20) \citet{Perley08};\,
      (21) \citet{Fynbo08b};\,
      (22) \citet{Jakobsson08c};\,
      (23) \citet{Jakobsson08b};\,
      (24) \citet{Vreeswijk08};\,
      (25) \citet{Thoene08b};\,
      (26) \citet{Thoene08};\,
      (27) \citet{Malesani08, Guidorzi09};\,
      (28) \citet{Jakobsson08};\,
      (29) \citet{Greiner09};\,
      (30) \citet{Ledoux07, Fox08};\,
      (31) \citet{Prochaska07b};\,
      (32) \citet{Prochaska07, Eliasdottir09};\,
      (33) \citet{Tanvir10};\,
      (34) \citet{Afonso08, Kruehler10};\,
      (35) \citet{Filgas08}.
      }
\end{table*}
\section{Methodology}

\label{sec:Methodology}

We moved through a 3D parameter space, consisting of mass ranges, metallicity upper limits and proportionality relations (i.e., $\delta$), in small step sizes and at each step compared the modelled LGRBR produced to our sample, using least $\chi^{2}$ statistics to asses the goodness of fit. The parameter details and methodology used are explained in the next sections.


\begin{table}
\caption{Summary of model parameters used in Eqn.~\ref{eqn:grb_rate}.}
  \centering
  \begin{tabular}{ l  l  l  l  l }
	  \hline
		  Parameter & Value or Range & Step size & Fixed \\
		  \hline
		  $\eta_{\mathrm{coll}}$ & $3.8\times10^{-3}$ & ... & y \\
		  $\eta_{time}$ & 0.14 (1) & ... & y \\
		  $\eta_{X-ray}$ & 0.57 \tablefootmark{a} & ... & y \\
		  $\eta_{redshift}$ & const.\tablefootmark{b} & ... & y \\
		  $\Delta \Omega$ & 0.077 & ... & y \\
		  $\Delta T$ & 3.5 & ... & y \\
		  $M_{\mathrm{BH}}/\mathrm{M_{\sun}}$ & 30.0 & ... & y \\
		  $M_{\mathrm{min}}/\mathrm{M_{\sun}}$ & 0.1 & ... & y \\
		  $M_{\mathrm{max}}/\mathrm{M_{\sun}}$ & 100 & ... & y \\
		  $\left(\mathcal{M}_{1}, \mathcal{M}_{2}\right)$ & $\left(7,12\right)$...$\left(7+\Delta\mathcal{M},12-\Delta\mathcal{M}\right)$ & $0.05$ & n \\
		  $\epsilon_{\mathrm{L}}/\mathrm{Z_{\sun}}$ & 0.1 - 1.8 & 0.05 & n \\
		  $\delta$ & 0 - 2.9 & 0.1 & n \\
		  $\eta_{other}$ & ... & ... & n \\
		  \hline\hline
   \end{tabular}
    \tablefoot{
      \tablefoottext{a}{Product of fraction of GRBs followed by XRT (2) and number of GRBs with an X-ray afterglow (3).}
      \tablefoottext{b}{Calculated based on the sample used, i.e., sub-sample size / full-sample size.}
      }
      \tablebib{
		(1) \citet{Greiner11};\
		(2) \url{http://heasarc.nasa.gov/docs/swift/archive/grb_table/stats/}
		(3) \url{http://www.mpe.mpg.de/~jcg/grbgen.html}
      }

\label{tab:Parameters}
\end{table}

\subsection{Fixed Parameters}
\label{Methodology:FixedParameters}

The parameters $\Delta T$, $\mathrm{\Delta \Omega}$ and $\eta_\mathrm{grb}$ in Eqn.~\ref{eqn:grb_rate} were assumed to be independent of redshift and luminosity. The values used can be found in Table~\ref{tab:Parameters}, but are discussed briefly in the following paragraphs.

The length of time of GROND observations (covered by our sample), $\mathrm{\Delta T}$, is 3.5 years (September 2007 - March 2010) and the solid angle of the BAT detector is $\mathrm{\Omega^{\it Swift}_{4\pi} = \frac{\Omega^{\it Swift}}{4\pi} = 0.11}$ \citep{Barthelmy05}. Therefore, for GROND, $\mathrm{\Omega^{\mathrm{GROND}}_{4\pi}} = \Omega^{\mathrm{La\,Silla}}_{4\pi}\cdot\Omega^{\it Swift}_{4\pi} = 0.077$ (i.e., the probability that the LGRB is observable to GROND after it has been detected by {\it Swift}).

The LGRB probability, $\eta_{\mathrm{GRB}}$, is related to $\eta_{\mathrm{BH}}$ and $\eta_{\mathrm{coll}}$. For the first of these parameters, see Eqn.~\ref{eqn:bheff}, we assume the mass above which a BH can form is taken to be $\mathrm{M_{BH}}\geq30\mathrm{M_{\sun}}$, over the possible mass range considered of $M_{\mathrm{min}}=0.1-M_{\mathrm{max}}=100\,\mathrm{M_{\sun}}$ in a Salpeter IMF, following the prescription used in \citet{NormanLanger06}.

The final parameter to be fixed is the collimation factor, $\eta_{\mathrm{coll}}$, being derived from the jet opening angle. The latter usually ranges between $1\degr$-$10\degr$ \citep[][]{Frail01, Cenko10} implying a collimation factor range of $1.5\times10^{-4} - 1.52\times10^{-2}$ \citep[utilising $\eta_{\mathrm{coll}} = 1 - \cos{\theta_{\mathrm{jet}}}$;][]{Frail01}. A median value of $\theta_{\mathrm{jet}} = 5\degr$, $\eta_{\mathrm{coll}} = 3.8\times10^{-3}$ is assumed throughout.

\subsection{Investigated Parameters}
\label{Methodology:InvestigatedParameters}

There are four parameters which were varied simultaneously to investigate which combination of these best agree with the observed LGRB rate. The parameters were: galaxy mass lower and upper limits $\left(\mathcal{M}_{1},\mathcal{M}_{2}\right)$ respectively, metallicity upper limit $\left(\epsilon_{L}\right)$, the proportionality power $\left(\delta\right)$ and the missing GRB probability, $\eta_{\mathrm{other}}$, all of which can be found in Eqn.~\ref{eqn:grb_rate}. As before, the parameter ranges investigated can be found in Table~\ref{tab:Parameters} and are briefly explained below.

Approximately 80\% of LGRB hosts are thought to lie within a mass range of $10^{9.4}\mathrm{M_{\sun}}$ to $10^{9.6}\mathrm{ M_{\sun}}$ \citep[][]{Savaglio09}. To test this hypothesis, a mass range boundary can be applied to the CSFH. This is done by moving the values of $M_{1}$ and $M_{2}$ to a central value, by equal step sizes, i.e., $\left(\mathcal{M}_{1},\mathcal{M}_{2}\right)$ = $\left(7,12\right)$...$\left(7+ \Delta \mathcal{M},12- \Delta \mathcal{M}\right)$. The initial boundaries are chosen to be close to the limits of that of the GOODS-Field survey and are the same as adopted by \citet[][]{Belczynski10}. The step size is set to $\Delta \mathcal{M}=0.05$.

The LGRB collapsar model requires a metallicity threshold to result in a LGRB. Also, LGRB host surveys show a similar preference for low-metallicity, as is indicated in Table~\ref{tab:grbhosts} where we give the median metallicities and masses of host galaxies from different surveys. Therefore, the metallicity upper limit, i.e., the galaxy metallicity below which the galaxy contributes to the CSFH, is set to range from $0.1\,-\,1.8\,\mathrm{Z_{\sun}}$ (the latter value being the limit at which the cuts stop affecting the CSFH in the model considered for the mass range chosen) with a step size of $\Delta Z=0.05~\mathrm{Z_{\sun}}$.

As mentioned previously, it is possible that there are other effects not considered by our model or one of the assumptions that we made may be inaccurate. This uncertainty is quantified by the parameter $\delta$. This black-box approach was carried out by \citet{Kistler09}, who found a value of $\delta=1.5$, which could be explained by metallicity cuts or an evolving luminosity function. As we have included metallicity effects and mass ranges, a positive value of $\delta$ would imply that (i) the LGRB LF is non-evolving, (ii) the stellar IMF is different to that considered in our model or (iii) the progenitor system is incorrect. To allow for a range of possibilities, $\delta$ is varied from 0 to 2.9 in steps of $\Delta \delta = 0.1$.

\begin{figure}
  \includegraphics[width=9cm]{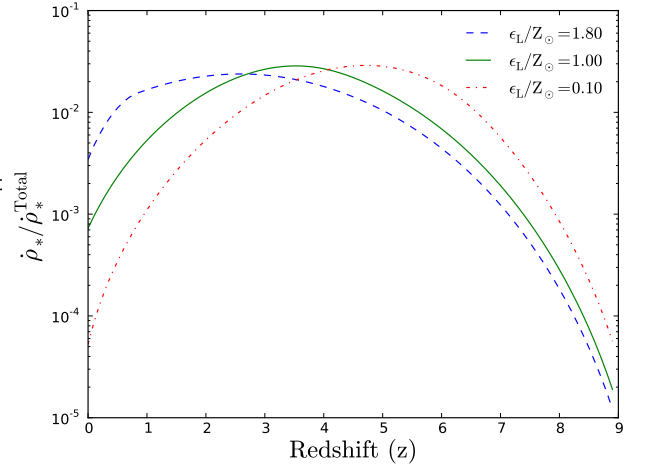}
  \caption{The resulting CSFH model when using different metallicity upper limits. Blue-dashed: $\epsilon_{\mathrm{L}}=1.8$, green-bold: $\epsilon_{\mathrm{L}}=1.0$ and red-dash-dotted: $\epsilon_{\mathrm{L}}=0.1$. Each curve utilises $\left(\mathcal{M}_{1},\mathcal{M}_{2}\right)=\left(7,12\right)$ and $\delta=0$. The normalisation, $\dot{\rho}_{*}^{Total}$, is the integrated CSFH from redshift $z=0-\infty$.}
  \label{fig:ParameterSpaces:a}
\end{figure}

\begin{figure}
  \includegraphics[width=9cm]{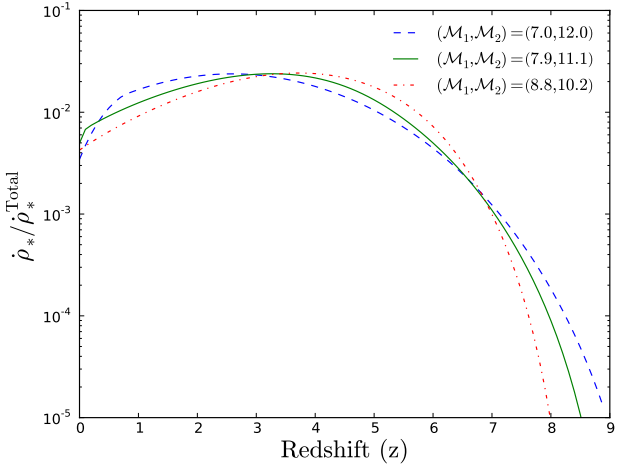}
  \caption{The resulting CSFH model when using different mass ranges. Blue-dashed: $\left(\mathcal{M}_{1},\mathcal{M}_{2}\right)=\left(7,12\right)$, green-bold: $\left(\mathcal{M}_{1},\mathcal{M}_{2}\right)=\left(7.9,11.1\right)$, red-dash-dotted: $\left(\mathcal{M}_{1},\mathcal{M}_{2}\right)=\left(8.8,10.2\right)$. Each curve utilises $\epsilon_{\mathrm{L}}=1.8$ and $\delta=0$. The normalisation, $\dot{\rho}_{*}^{Total}$, is the integrated CSFH from redshift $z=0-\infty$.}
  \label{fig:ParameterSpaces:b}
\end{figure}

\begin{figure}
  \includegraphics[width=9cm]{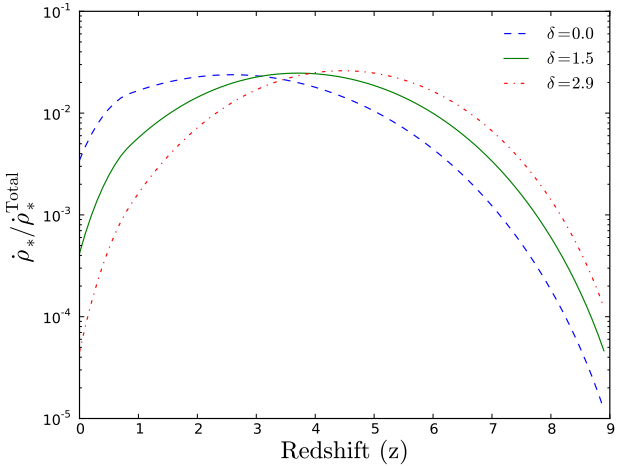}
  \caption{The resulting CSFH model when using different power dependences. Blue-dashed: $\delta=0$, green-bold: $\delta=1.5$, red-dash-dotted: $\delta=2.9$. Each curve utilises $\epsilon_{\mathrm{L}}=1.8$, $\left(\mathcal{M}_{1},\mathcal{M}_{2}\right)=\left(7,12\right)$. The normalisation, $\dot{\rho}_{*}^{Total}$, is the integrated CSFH from redshift $z=0-\infty$.}
  \label{fig:ParameterSpaces:c}
\end{figure}

The final parameter we consider free is $\eta_{\mathrm{other}}$, obtained from Eqn.~\ref{eqn:grb_rate}. As a result of its unknown nature, it will be determined from the model fitting that is described in Sect.~\ref{Methodology:Modelling}. Unlike the static parameters described in the Sect.~\ref{Methodology:FixedParameters}, the parameters described in this section are redshift and mass dependent. Different combinations of these parameters can produce similar outcomes, i.e., a degeneracy and each parameter can have an effect on the other. For example, when the mass range is brought inwards, for each redshift, it will limit also the metallicities, as galaxies grow they will produce more metals. The only way to see which combination is preferred is to look for the best-fit to the data in a systematic approach. The effect of changing each parameter individually is depicted in Figs.~\ref{fig:ParameterSpaces:a}, ~\ref{fig:ParameterSpaces:b} and ~\ref{fig:ParameterSpaces:c}.

\begin{table}
    \centering
    \caption{Median metallicities and masses for different LGRB host galaxy programs.}
    \label{tab:grbhosts}
    \begin{tabular}{ l  l  l  l  l }
	    \hline
		    Survey & $Z$ & $\delta_{Z}$\tablefootmark{a} & $\log_{10}\left( M_{*}/\mathrm{M_{\sun}}\right)$ & $\delta_{\mathcal{M}}$\tablefootmark{a} \\ \hline
		    & ($\mathrm{Z_{\sun}}$) & dex &  & dex \\
		    \hline
		    \citet{Sven10} & 0.54 & 0.65 & 9.12 & 1.84 \\
		    \citet{Savaglio05} & 0.26 & 0.12 & 9.32 & 0.75 \\
		    \citet{Mannucci11} & 0.61 & 0.30 & 9.33 & 0.64 \\
		    \citet{Rau10}\tablefootmark{b} & 0.13 & 0.62 & ... & ... \\
		    \citet{Kruehler11} & ... & ... & 9.76 & 0.38 \\
	    \hline\hline

    \end{tabular}
    \tablefoot{
      \tablefoottext{a}{Standard deviation.}
      \tablefoottext{b}{Damped Lyman-$\alpha$ systems at $z\sim2$.}
      }
\end{table}

\subsection{Modelling}
\label{Methodology:Modelling}

The 3D parameter space is to be investigated in a brute-force approach, in the following steps:
\begin{enumerate}
  \item The selected samples are first binned in log-luminosity space with a bin size of $\log\left(L/\mathrm{erg\,s^{-1}}\right)=0.5$ for the spectroscopic and photometric samples, and $1.0$ for the upper limit 1 and upper limit 2 samples (see e.g., Fig.~\ref{fig:GRDB_LuminosityRedshiftSample:b}). 
  \item A non-evolving LGRB LF (Eqn.~\ref{eqn:GRBLF}) is then fit to the luminosity distribution (corresponding to each sample) using a least squares algorithm taken from the {\it SciPy}\footnote{\url{http://www.scipy.org/}} library.
  \item The chosen sample is then binned in redshift space (bin sizes; spectroscopic: 1.74, photometric: 1.87, upper limit 1: 1.6, upper limit 2: 2.0) where bin sizes are chosen to ensure a limited number of bins ($>80$\%) have zero counts.
  \item The LGRB number density model (Eqn.~\ref{eqn:grb_rate}) is then calculated utilising a given $\left(M_{1},M_{2}\right)$, $\delta$, $\epsilon_{\mathrm{L}}$ and LGRB LF determined in step 2, using the same redshift bin sizes as in the previous step, i.e., $N\left(z_{1},z_{2}\right)$.
  \item $\eta_{\mathrm{other}}$ is calculated by taking the median value of the ratio of the binned data to the expected model data (a median is favoured as the high-z bins contain low counts and can dominate the slope of the best-fit).
  \item The data is then compared to the model predictions using a least $\chi^{2}$-test $\left(\chi^{2}=\sum\frac{(x_{\mathrm{expected}}-x_{\mathrm{observed}})^{2}}{\delta x_{\mathrm{observed}}}\right)$. The count errors are assumed to be Poisson distributed.
  \item This is then repeated for all of the parameters in the 3 dimensional space described in Table~\ref{tab:Parameters}.
\end{enumerate}

\begin{figure}
  \includegraphics[width=9cm]{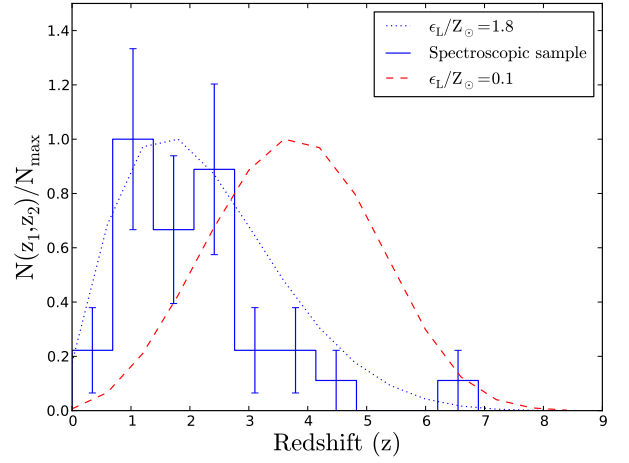}
  \caption{Examples of LGRBR models in comparison to the spectroscopic sample. The blue-solid line is spectroscopic sample, the blue-dashed is the model with input parameters $\delta=0$, $\epsilon_{\mathrm{L}}=1.8$, $\left(\mathcal{M}_{1},\mathcal{M}_{2}\right)=\left(7, 12\right)$ and the red-dashed line is the same but with a metallicity limit of $\epsilon_{\mathrm{L}}=0.1$. All three histograms are normalised to their peak value to highlight their main differences.}
\label{fig:Results:GRBRateFit}
\end{figure}

\noindent An example of the binned data and the corresponding LGRB rate models for two metallicity constraints can be seen in Fig~\ref{fig:Results:GRBRateFit}.

\subsection{Summary of Assumptions}
To summarise, the assumptions made throughout this section are:

\begin{itemize}
 \item LGRBs produce an X-ray afterglow and are collimated.
 \item LGRBs are formed via the single progenitor collapsar model from stars above a mass of $30\,\mathrm{M_{\sun}}$.
 \item A Salpeter IMF between $0.1-100\,\mathrm{M_{\sun}}$, is assumed.
 \item LGRBs have a static normal (Gaussian) luminosity function.
 \item Galaxies obey a redshift-evolving mass function.
 \item Galaxies lie on the mass-metallicity relation.
 \item Downsizing is described by an evolving quenching mass (mass upper limit).
\end{itemize}

\section{Results}
\label{sec:Results}
\begin{figure*}

    \subfloat[Spectroscopic (S) Sample, $\df=3$.]{\label{fig:Results:ChiSquareBestFits:Spectroscopic}\includegraphics[width=10cm]{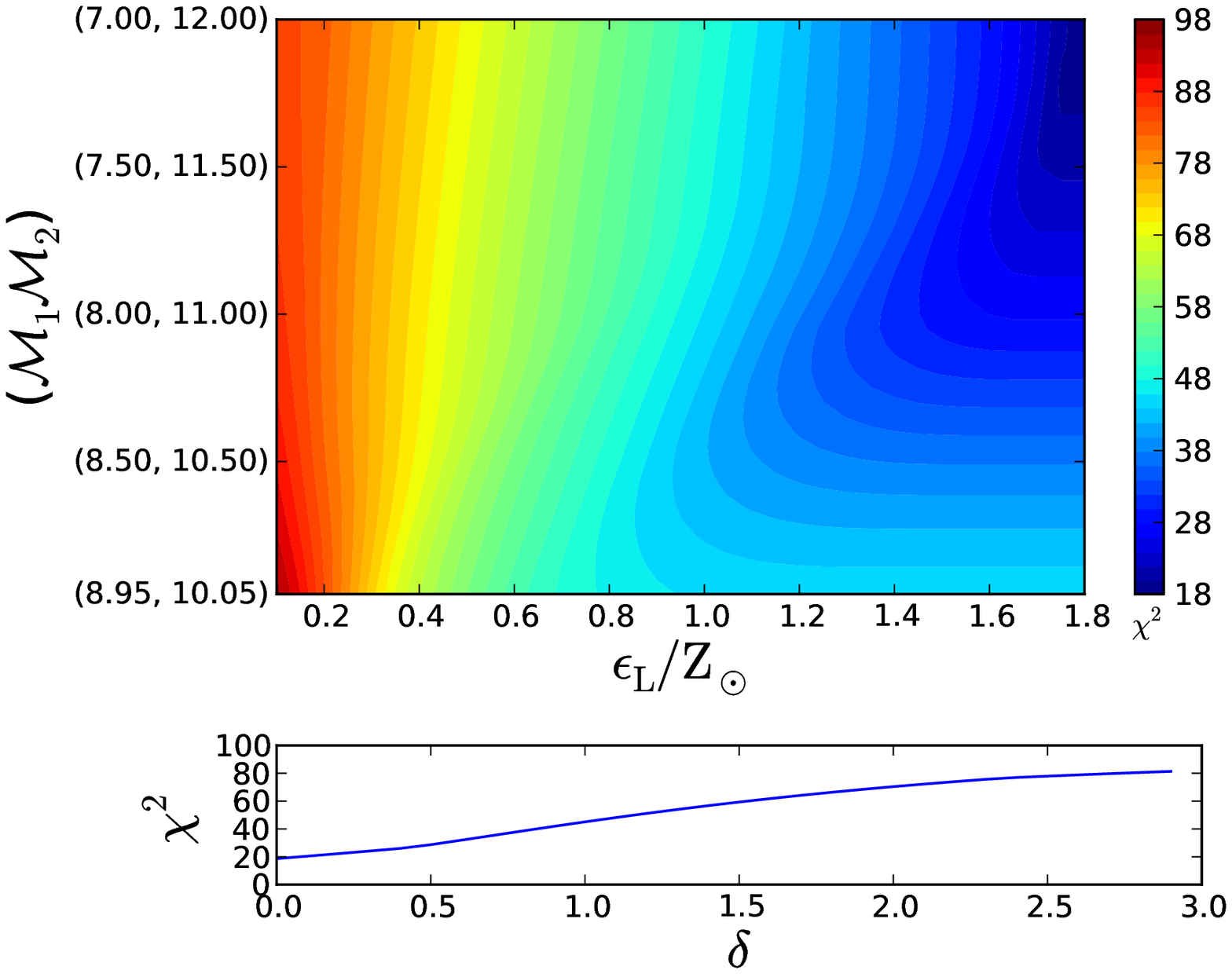}}
    \subfloat[Photometric (P) Sample, $\df=4$.]{\label{fig:Results:ChiSquareBestFits:Photometric}\includegraphics[width=10cm]{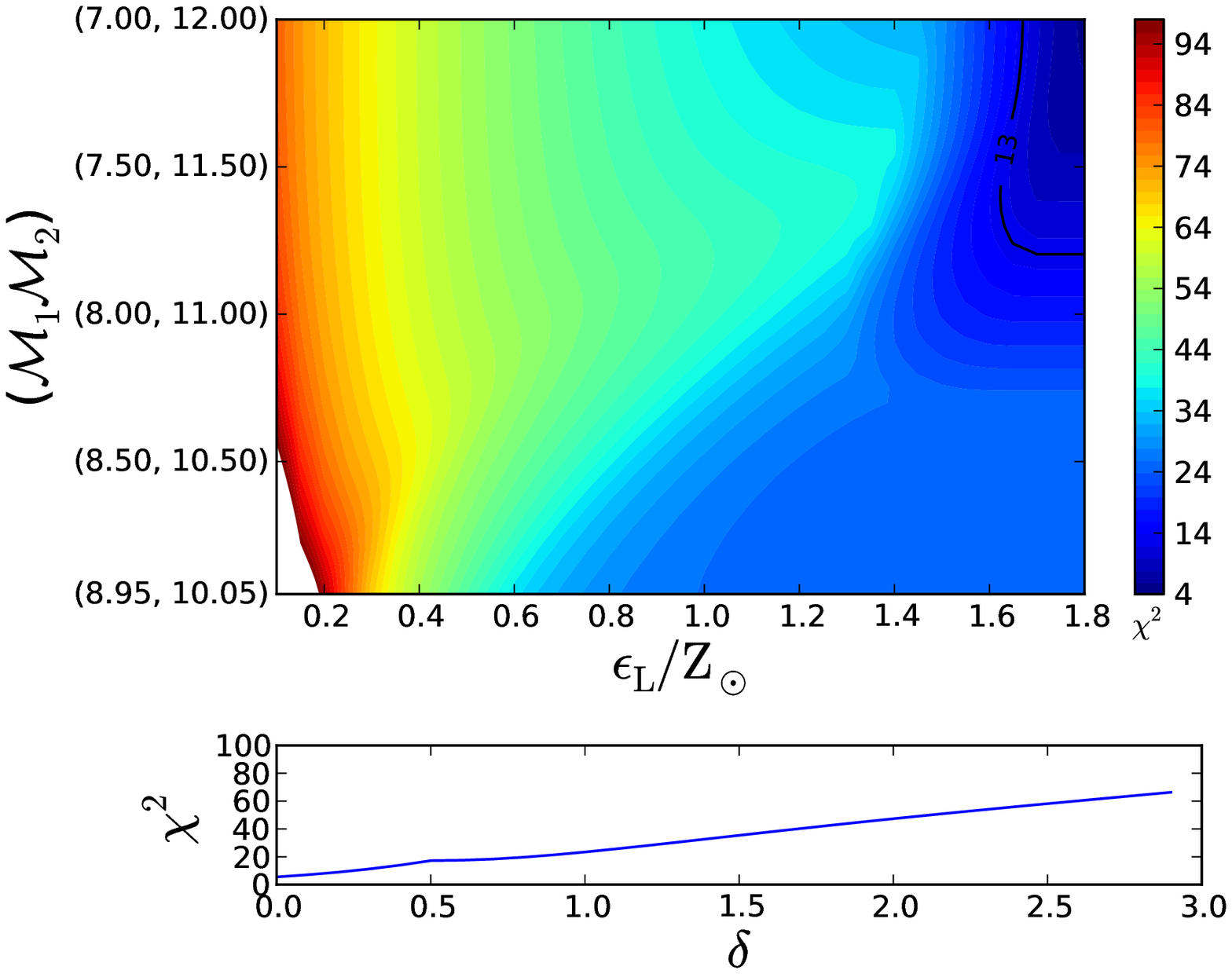}}

    \subfloat[Upper limit 1 (UL1) Sample, $\df=5$.]{\label{fig:Results:ChiSquareBestFits:UpperLimit1}\includegraphics[width=10cm]{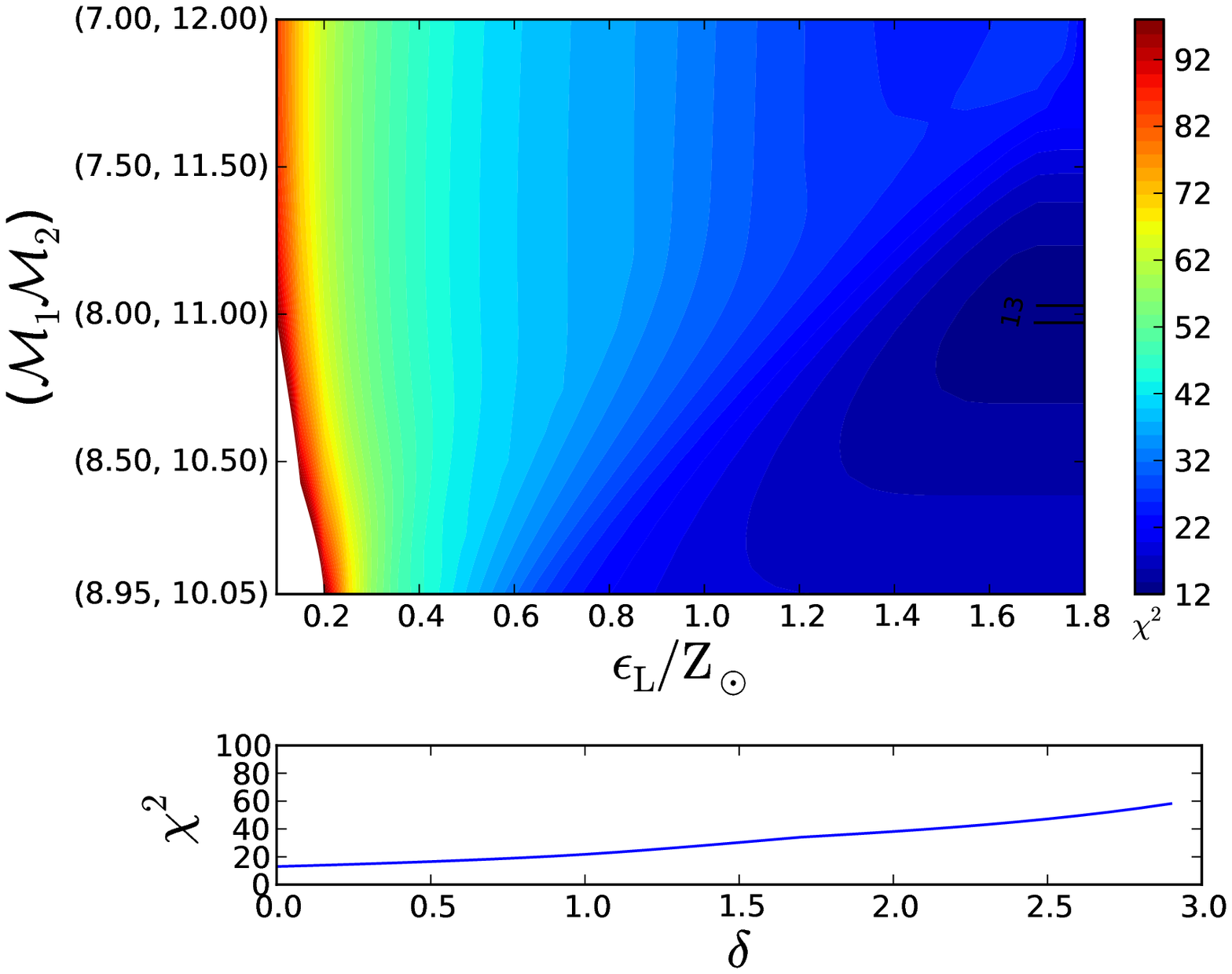}}
    \subfloat[Upper limit 2 (UL2) Sample, $\df=5$.]{\label{fig:Results:ChiSquareBestFits:UpperLimit2}\includegraphics[width=10cm]{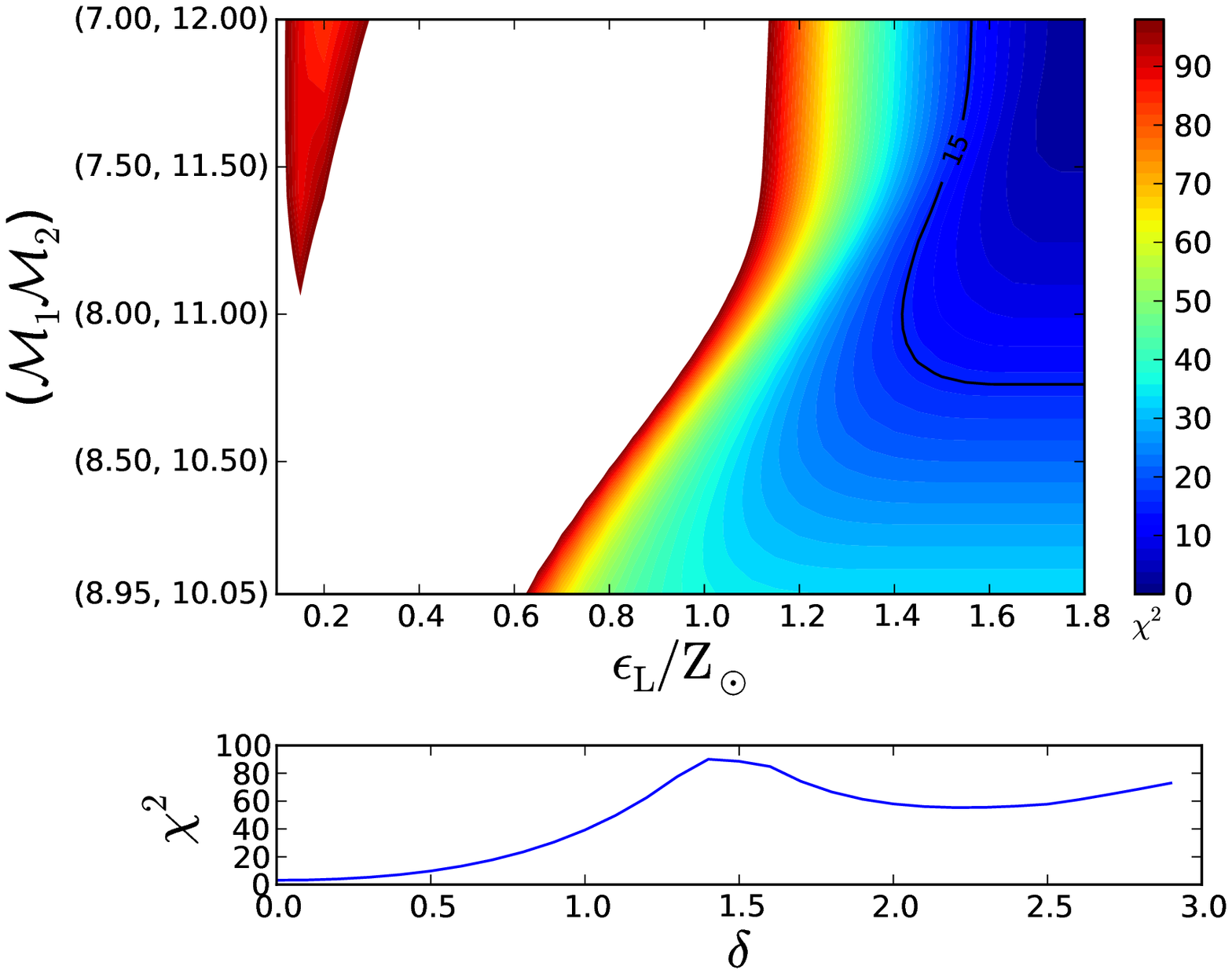}}

   \caption{The best-fit plots for each sample investigated. \textbf{Top Panels}: The $\chi^{2}$ contour plot for $\delta=0$, white areas denote $\chi^{2}$ values much larger than the colour scale shown. Only the contour of $\delta=0$ is displayed as the fits begin to get worse when $\delta$ starts to increase. The black lines denote the maximum $\chi^{2}$ to be an acceptable fit to the $1\%$ level. \textbf{Bottom Panels}: The $\chi^{2}$ values for all of the best-fits in $\delta$ space, depicting the worst fits for progressively increasing $\delta$.}
\end{figure*}

Given the methodology outlined in Sect.~\ref{sec:Methodology}, a LGRB number distribution was generated utilising a 3D parameter space consisting of metallicity limits, mass ranges and missing redshift effects. These models were then compared to the different samples described in Sect.~\ref{subsec:GrondSamplesDarkBurstsSamples}, of varying completeness levels, and the best-fit results were determined using a least $\chi^{2}$ test.

\subsection{Parameter Spaces}
\label{sec:Results:ParameterSpaces}
The results of the fits can be seen in Figs.~\ref{fig:Results:ChiSquareBestFits:Spectroscopic}, \ref{fig:Results:ChiSquareBestFits:Photometric}, \ref{fig:Results:ChiSquareBestFits:UpperLimit1} and \ref{fig:Results:ChiSquareBestFits:UpperLimit2}. Each figure depicts the contour plot for metallicity vs mass of the best-fit values for $\delta=0$ (upper panels). For all of the distributions investigated, the best-fit $\delta$ was zero and so only a single contour plot is displayed. For clarity to the reader, the best-fit $\chi^{2}$ for all $\delta$ values considered is shown (see lower panels). All the best-fit values can be found in Table~\ref{tab:Results:ModifiedCSFR}. The zero value of $\delta$ implies that there are no missing redshift effects in the modelling that has been used (this is discussed further in Sect.~\ref{sec:Discussion}).

The four samples, S, P, UL1 and UL2 show a preference for a CSFH-LGRBR connection with no strong global metallicity restrictions. The S and P samples show that no mass range limitations for the host galaxies are preferred. The UL1 and UL2 samples show a preference for the metallicity limits $\epsilon_{\mathrm{L}}=1.7\,\mathrm{Z_{\sun}}$ and $1.75\,\mathrm{Z_{\sun}}$ and the mass ranges $\left(\mathcal{M}_{1},\mathcal{M}_{2}\right)=\left(8.0,11.0\right)$ and $(7.15,11.85)$ respectively. These limits are very lax in comparison to the normal measured results of $\epsilon_{\mathrm{L}}\leq0.3\,\mathrm{Z_{\sun}}$ \citep[see e.g.,][]{NormanLanger06,Salvaterra07,Campisi10,Virgili11} and the range of $\left(\mathcal{M}_{1},\mathcal{M}_{2}\right)=(9.4,9.6)$ \citep{Savaglio09}, and are interpreted as no limit. The predominant difference in our best-fit model compared to previous studies in this area is a result of the differences in the peak of the redshift number distribution whereby previous samples usually peaked at $z\sim4$, and required a metallicity cut-off (see Fig.~\ref{fig:Results:GRBRateFit}). As we are comparing this to an experimental data set, the way in which a sample is selected is very important (this is discussed in more detail in the next section).
We note that there is no acceptable fit to the S sample to the $1\%$ level, but is included for completeness. Also, there are models present in the contour plot that are still acceptable fits at the $1\%$ level for the following metallicity ranges and mass boundaries: $\left(\mathcal{M}_{1},\mathcal{M}_{2}\right)=(7,12)-(7.8,11.2)$ and $\epsilon_{\mathrm{L}}/\mathrm{Z_{\sun}}=1.65-1.8$, $\left(\mathcal{M}_{1},\mathcal{M}_{2}\right)=(7.95,11.05)-(8.05,10.95)$ and $\epsilon_{\mathrm{L}}/\mathrm{Z_{\sun}}=1.65-1.8$, $\left(\mathcal{M}_{1},\mathcal{M}_{2}\right)=(7,12)-(8.25,10.75)$ and $\epsilon_{\mathrm{L}}/\mathrm{Z_{\sun}}=1.45-1.8$, for the P, UL1 and UL2 samples respectively ($1\%$ acceptance boundaries are depicted as bold-black lines in Figs.\ref{fig:Results:ChiSquareBestFits:Photometric}, \ref{fig:Results:ChiSquareBestFits:UpperLimit1} and \ref{fig:Results:ChiSquareBestFits:UpperLimit2}).

\begin{table*}
\caption{Best fit properties of all the redshift samples.}
\centering
 \begin{tabular}{ l l l l l l l l l l }
  \hline
  Sample & $\epsilon_{\mathrm{L}}$ & $\mathcal{M}_{1}$ & $\mathcal{M}_{2}$ & $\delta$ & $\eta_{\mathrm{other}}$~\tablefootmark{*} & $\chi^{2}/\df$ & $a$ & $\chi^{2}_{\mathrm{modified}}/\df$ & $\eta_{\mathrm{other}}^{\mathrm{modified}}$~\tablefootmark{*}\\
  \hline
  & $\mathrm{Z_{\sun}}$ & $\log_{10}\left(\mathrm{M/M_{\sun}}\right)$ & $\log_{10}\left(\mathrm{M/M_{\sun}}\right)$ & & $\times10^{3}$ & & $\mathrm{\sfrunit}$ & & $\times10^{3}$ \\
  \hline

  S & 1.80 & 7.00 & 12.00 & 0.0 & $2.5\pm1.7$ & 18.65/3 & 0.063 & 18.49/3 & $2.5\pm1.7$\\
  P & 1.80 & 7.00 & 12.00 & 0.0 & $3.6\pm2.6$ & 4.81/4 & 0.050 & 6.92/4 & $3.6\pm0.7$\\
  UL1 & 1.70 & 8.00 & 11.00 & 0.0 & $7.0\pm3.2$ & 12.98/5 & 0.013 & 12.86/5 & $6.3\pm1.7$ \\
  UL2 & 1.75 & 7.15 & 11.85 & 0.0 & $4.1\pm1.9$ & 2.19/5 & 0.032 & 4.12/5 & $4.1\pm0.5$ \\
  \hline\hline
 \end{tabular}
 \label{tab:Results:ModifiedCSFR}
 \tablefoot{
  \tablefoottext{*}{Errors are the standard deviation of the first 3 bins.}
 }  
\end{table*}

\subsection{Completeness Levels}
\label{Results:OtherSamples}

\begin{figure}
  \includegraphics[width=9cm]{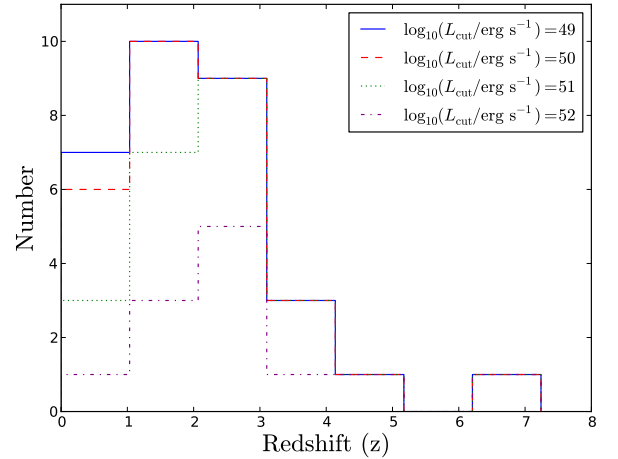}
  \caption{LGRB redshift number distribution. Each distribution depicts the sample after LGRBs below a specific luminosity limit are removed. It can be seen the redshift peak remains in the range 1-3.}
  \label{fig:Results:GRDBLumCut}
\end{figure}

The differences observed between these results and that of previous work can be explained by the redshift completeness of the samples that have been considered. To generate samples of complete redshift, the common approach is to take LGRBs that are above a specific luminosity value (luminosity cut). Such a method assumes that for all LGRBs above the chosen luminosity cut, the probability of measuring the LGRB redshift is equal and independent of redshift. A luminosity cut in itself is a perfectly acceptable thing to do as one can implement the same choices in the model that is used by placing the same luminosity cuts in Eqn.~\ref{eqn:GRBLF}. However, as it has been mentioned previously, redshift follow-up is not a consistent process. Ground-based programs are usually biased towards following potentially interesting GRBs (i.e., at high-z), and measurements are easier to attain in the case of the more luminous, quickly detected bursts with smaller telescopes. There remains a large gap, both from the spectroscopic desert in the redshift range $z=1-2$ \citep[see e.g.,][]{Steidel04}, but also in the mid-range redshift $z=2-3$ as the LGRB is not visible to the smaller telescopes and the time is not always allocated for such bursts on the large telescopes. Over the past few years, this has begun to change and all types of LGRB are being targeted for follow-up, leading to highly complete samples, much like the one in this paper. For a LGRBR that is biased to galaxies with an $\epsilon_{\mathrm{L}}=0.1$, the LGRB number density would need to peak at redshift 4 (see Fig.~\ref{fig:Results:GRBRateFit}). Placing luminosity cuts on the sample in this paper, it can be seen that the redshift peak remains in the range 1-3 (see Fig.~\ref{fig:Results:GRDBLumCut}) and even very large cuts cannot place it in the redshift bin of 4. We also note that for the S sample no acceptable fit was found, to the $1\%$ level, to fulfil a null hypothesis, in comparison to the other three samples. This is in agreement with the argument presented, as $2/3$ photometric redshifts and 1 upper limit lie in the redshift $z\sim3$, which are excluded from the spectroscopic sample.

\section{Discussion}
\label{sec:Discussion}
The possible connection between the CSFH and LGRBR was investigated utilising a model of the CSFH and a 3D parameter space of: mass ranges, metallicity limits and proportionality laws. This was used in combination with a highly complete LGRB redshift measured sample.
\subsection{LGRB Probability}
\label{subsec:Discussion:CSFR-GRBRApplications}
\label{subsec:Discussion:GRB}
To calculate the LGRB probability parameter, $\eta_{\mathrm{grb}}$ (Eqn.~\ref{eqn:grbprob}), we made a priori assumptions about several values (e.g., black hole mass and jet collimation). However, the parameters taken from the literature that influence the probability of a star forming a LGRB have a possible range of acceptable values (see Sect.~\ref{Methodology:FixedParameters}). 
Due to the inclusion in our calculation of $\eta_{\mathrm{grb}}$ of a normalisation parameter, $\eta_\mathrm{other}$, a change in the physical parameters $\eta_\mathrm{coll}$ (and thus $\theta_{\mathrm{coll}}$) and $\eta_\mathrm{BH}$ (and thus $M_{\mathrm{BH}}$) would not change the shape of the resulting LGRBR distribution, since the value $\eta_{\mathrm{grb}}$ would counteract changes in $\eta_{\mathrm{coll}}$ and $\eta_{\mathrm{BH}}$. Assuming that we have included all the LGRB probabilities and thus $\eta_{\mathrm{other}}=1$, the resulting changes in $\theta_{\mathrm{coll}}$ and $M_{\mathrm{BH}}$ can be quantified by:

\begin{equation}
  M_{\mathrm{BH}}' = \left[\frac{-(\alpha+1)}{\eta_{\mathrm{other}}}\int_{M_{\mathrm{BH}}}^{M_{\mathrm{max}}}\psi\left(m\right)\mathrm{d}m + M^{-(\alpha+1)}\right]^{\frac{1}{\left(\alpha+1\right)}}
  \label{eqn:eta_sn_prime}
\end{equation}

\begin{equation}
 \theta_{\mathrm{coll}}' = \cos^{-1}\left(1-\frac{\eta_{\mathrm{coll}}}{\eta_{\mathrm{other}}}\right).
 \label{eqn:eta_beam_prime}
\end{equation}

\noindent The rescaled values for all the samples investigated can be found in Table~\ref{tab:Discussion:Predictions} and show two extreme cases: (i) large progenitor masses and (ii) small jet opening angles, for which the first property relies on the LGRB collapsar mechanism. The values of (i) and (ii) were initially chosen to be $\mathrm{M_{BH}}=30\mathrm{M_{\sun}}$ and $\theta_{\mathrm{jet}}=5\degr$ respectively (see Sect.~\ref{sec:Methodology}). For a LGRB to occur, we required the formation of a BH which we assume forms a progenitor above a specific mass, $M_{\mathrm{BH}}$. Such large masses of the progenitor imply that LGRBs form from direct collapse to a BH rather than SN fall back under the collapsar model \citep[see e.g.,][]{Heger03}.  Smaller jet opening angles of $\sim1\degr$, as mentioned previously, are possible but would have effects on, for example, the derived collimated energies of LGRBs \citep[see e.g.,][]{Racusin09}.

Despite the fact that we have considered these parameters separately, it is also equally valid to tune each value at once to give the same final results. To improve upon this, each probability parameter must be measured to more precision before the CSFH can be used to constrain any single one of them. Finally, it is also possible that one of these parameters could also evolve with redshift, but as $\delta=0$ was shown to be the best-fit in Sect.~\ref{sec:Results}, this again is unlikely to be the case.
\begin{table*}
\caption{Redshift and LGRB probability predictions.}
\centering
 \begin{tabular}{ l l l l l l l }
  \hline
  Sample & $5<z<6$ & $6<z<8$ & $z>8$ & $\theta_{\mathrm{coll}}'$\tablefootmark{*} & $M_{\mathrm{BH}}$ \tablefootmark{*}\\
  \hline
  & \% & \% & \% & $\degr$ & $\mathrm{M_{\sun}}$ \\
  \hline

  S\tablefootmark{N} & 3.7 & 1.2 & 0.0169 & $0.98\pm0.36$ & $99.88\pm36.78$ \\
  S\tablefootmark{M} & 2.7 & 0.1 & 0.0002 & $0.98\pm0.36$ &  $99.88\pm36.78$ \\
  P\tablefootmark{N} & 3.7 & 1.2 & 0.0169 & $0.82\pm0.30$ & $99.91\pm36.76$ \\
  P\tablefootmark{M} & 5.6 & 3.4 & 0.0030 & $0.82\pm0.30$ & $99.91\pm36.76$ \\
  UL1\tablefootmark{N} & 4.9 & 1.3 & 0.0050 & $0.58\pm0.21$ & $99.95\pm36.77$ \\
  UL1\tablefootmark{M} & 7.8 & 9.3 & 8.0831 & $0.60\pm0.22$ & $99.95\pm36.77$\\
  UL2\tablefootmark{N} & 3.7 & 1.2 & 0.0147 & $0.75\pm0.28$ & $99.93\pm36.76$ \\
  UL2\tablefootmark{M} & 6.9 & 6.9 & 2.3077 & $0.75\pm0.28$ & $99.93\pm36.76$ \\

  \hline\hline
 \end{tabular}
 \label{tab:Discussion:Predictions}
\tablefoot{
\tablefoottext{N}{ Normal CSFH model (Eqn.~\ref{eqn:csfr}) using the best-fit properties given in Table~\ref{tab:Results:ModifiedCSFR}.}
\tablefoottext{M}{Modified CSFH model (Eqn~\ref{eqn:modifiedCSFR}) using the best-fit properties given in Table~\ref{tab:Results:ModifiedCSFR}.}
\tablefoottext{*}{Uncertainties are given as $1\sigma$ deviations, as the propagated errors of Eqn.~\ref{eqn:eta_sn_prime} and Eqn.~\ref{eqn:eta_beam_prime} are dominated by $1/\eta_{\mathrm{other}}^{2}$ and are thus underestimated.}
}
\end{table*}

\subsection{High-z Predictions of the CSFH}
\label{Results:High-zCSFR}
\begin{figure}

\includegraphics[width=9cm]{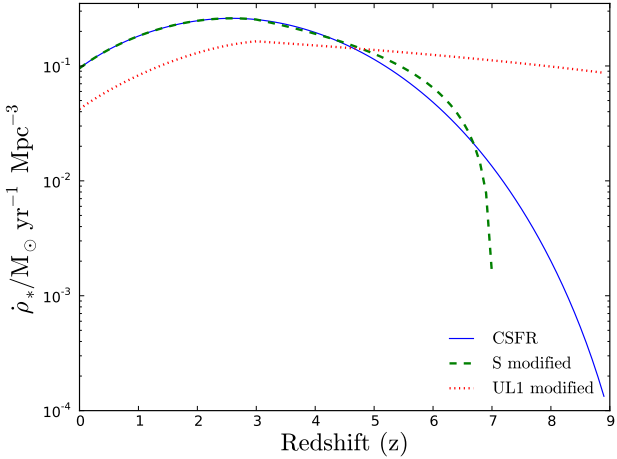}
 \caption{The bold-blue line is the CSFH model described by eqn~(\ref{eqn:csfr}). The green-dashed and the red-dotted lines depict the best-fit modified CSFH for the S and UL1 distributions respectively.}

\label{fig:Results:ModifiedCSFR}
\end{figure}

Galaxy detection drops off very quickly with higher redshifts due to instrumental limitations making LGRBs complementary probes to studying high-z star formation, provided we understand the relation between the LGRBR and the CSFH. As a result of the empirically determined models outlined in Sect. \ref{sec:CosmicStarFormationAndGRBRate} being limited to redshifts $z<3$ and the possibility that the CSFH flattens out at at high redshift, we modify the CSFH model to a linear function for redshifts of $z>3$ \citep[similar to][]{Daigne06}:

 \begin{equation}
   \dot{\rho}\left(z\right) = \left\{
     \begin{array}{l l}
 	\dot{\rho}\left(z\right) & \quad \mathrm{if}\,z\le3 \\
	\dot{\rho}\left(z=3\right) - az & \quad \mathrm{if}\,z>3 \\

     \end{array} 
   \right.,
 \label{eqn:modifiedCSFR}
 \end{equation} where $a$ is a constant to be determined. The parameter $a$ is set to vary from the slope of the CSFH at $z=3$ and is increased to a flat distribution over the range $\log_{10}\left(a/\sfrunit\right)=0.4-2.5$. Each parameter fit was then compared to each LGRB sample and the best-fit $a$ can be found in Table~\ref{tab:Results:ModifiedCSFR} and the resulting change in the best-fit CSFH can be seen in Fig.~\ref{fig:Results:ModifiedCSFR}.

All of the samples show no preference for a CSFH flattening at high redshift except the upper limit 1 distribution, which shows a preference for a CSFH that remains constant with evolving redshift. However, the $\Delta \chi^{2}$ between the modified and unmodified CSFH is 0.12, and so both are still feasible solutions. Overall, the other three samples show no preference for a flattening in the CSFH at higher redshifts. This is in contradiction to the work by \citet{Kistler09}, who showed that CSFH would be higher than thought, utilising the {\it Swift} LGRB sample and a single high-z LGRB. As discussed in Sect.~\ref{Results:OtherSamples}, this is understood as being a result of the extended efforts of measuring redshifts of LGRBs for $z>6$. Also, work carried out by \citet{Daigne06} show that they required a CSFH that increased with redshift. However, this would imply higher star formation rates at the early stages of the Universe, which are not seen observationally or with simulations. Due to this unphysical nature, they propose that the LGRB mechanism is evolving with redshift. Again, due to the preference of $\delta=0$ in our studies, our analysis suggests that there is no evolution of the LGRB mechanism required.

Given the best-fit models the total fraction of LGRBs that exist in each redshift range is predicted for each distribution analysed (values can be found in Table~\ref{tab:Discussion:Predictions}). A fraction of $\sim1.2\%$ of the LGRB population existing $z>6$ is calculated, similar to the value of $\sim1\%$ from \citep{Campisi10}, within the error of \citet{Greiner11} who predict $5.5\pm2.8\%$ for $z>5$ and consistent with other works \citep[][]{Perley09, Fynbo09}. While this sounds like a small fraction, it is much larger than the corresponding fractions for AGN/QSO \citep{Willott10}.

\subsection{Summary of Results and Limitations}
\label{subsec:EmpiricalLawsAssumptionsDegeneracy}
It had been previously thought that the LGRB host sample was biased to a specific range of masses with low metallicity. We find acceptable fits for a mass range of $10^{7}-10^{12}\mathrm{M_{\sun}}$, which would suggest that the LGRB rate is a sensitive measure of the faint- and massive-end of the mass function of galaxies. We note that the proportions of galaxies are determined by the specified GMF and the quenching mass, both described in Sect.~\ref{sec:CosmicStarFormationAndGRBRate}. The high mass range is in good agreement with \citet{Kruehler11}, who show that by selecting hosts based on high extinction more massive galaxies were found. On the contrary, the GMF would imply there are more missing low-mass galaxies and that LGRBs are selecting star forming regions, with no biases.  Secondly, the results of Sect.~\ref{sec:Results:ParameterSpaces} also show that there is no strong, $\sim0.1-0.3\,\mathrm{Z_{\sun}}$ metallicity preference for the host galaxy at the 95\% redshift-complete level. It should be noted that this is a global metallicity and does not reflect directly on the LGRB model itself, but on the properties of the LGRB hosts as a whole. Other types of cuts have been implemented where metallicity dispersion is also considered \citep[][]{Niino11a, Wolf07}, however, as our model solution requires no cuts at all, this would not affect the outcome. Thirdly, the $\delta$ parameter chosen to quantify any effects not considered within our model favours a value of zero. Such a result implies the following: (i) the stellar IMF need not be evolving, (ii) the LGRB LF is also redshift independent and (iii) the progenitor system is consistent throughout all redshifts considered.

In summary, the results show that LGRBs are occurring in any type of galaxy (the galaxy number density obeying an evolving GMF of full mass range), with no strong metallicity limits. The only requirement is that there is star formation occurring, in agreement with simulations that suggest LGRB host galaxies preferentially have high specific star formation rates \citep[e.g.,][]{Courty07, Mannucci11} and observational evidence that LGRBs primarily occur in regions of high star formation \citep{Fruchter06}.

As with any model there are limitations. For low redshifts ($z<3$) we had used empirically calibrated models, that allowed the freedom of their parameters to be removed (see e.g., Eqns.~\ref{eqn:sfr},~\ref{eqn:gmf},~\ref{eqn:metallicity} or~\ref{eqn:quenchingmass}). To make reliable judgements at high redshifts ($z>3$), further observations able to study these relations at higher redshift are needed or other techniques used such as, Monte Carlo Markov Chain \citep[e.g.,][]{Virgili11}, Principal Component Analysis \citep[e.g.,][]{Ishida11}, or simulations introduced into our model \citep[e.g.,][]{Campisi10, deSouza11, Ishida11}. Another important consideration is the assumptions and simplifications made at the beginning of the modelling, i.e., the stellar IMF, the LGRB luminosity function, the LGRB explosion mechanism and the GMF. Many of the parameters kept constant could also be left free.

The form of the stellar IMF, at present, is still a lively debated issue and could be modified by changing the slope ($\alpha$), the change over short time scales and also the change over redshift (top-heavy; see Sect.~\ref{subsec:InitialMassFunction}). Any of these changes would result in a modification of the following models that use an IMF for their determination: SFR (Sect.~\ref{subsec:SFR}), GMF (Sect.~\ref{subsec:GalaxyMassFunction}) and the LGRBR probability (Sect.~\ref{subsec:GRBLuminosityFunction}). For example, \citet{Wang11} show that the CSFH-LGRBR connection requires no constraints if a redshift dependent stellar IMF is used. For our given framework we require no evolution of the IMF, however, any big changes to the form of the stellar IMF would require deeper analysis.

Secondly, the luminosity function of the LGRB was assumed to take the form of a normal (Gaussian function). There are many other forms, as mentioned previously, such as Schechter functions, log-normals and redshift dependent functions that could be utilised. Again, each one implies different physics, and will naturally influence the final result.

The minimum mass of a star to form a BH was set to $M_{\mathrm{BH}}\geq30\,\mathrm{M_{\sun}}$, as the collapsar model was assumed for LGRB creation. However, there is a range of possible lower range masses \citep{Nomoto10} and also secondary mechanisms for BHs to generate a LGRB, for example SN-fallback \citep[for a review see][]{Fryer07}. There are also different types of progenitor systems (other than a WR) that are thought to be possible to generate a LGRB, e.g., NS-White Dwarf binaries~\citep{Thompson09}, Helium-Helium binaries~\citep{Fryer05}, Quark stars~\citep{Ouyed05}, Be/Oe stars~\citep{Martayan10, Eldridge11}, blue-stragglers~\citep{Woosley06} and red-giants~\citep{Eldridge11}. Different mechanisms have been investigated before, but not in combination with the parameters of this paper \citep[see][]{Young07}. 

Finally, the primary difference of the studies of this paper is the redshift completeness of the distribution. The normal way of improving completeness levels is to choose LGRBs above a specific luminosity to compensate for the limitations of the detector at increasing redshift. However, redshift measurements are not only dependent on the brightness of the afterglow, and follow-up of GRBs is not always consistent for many reasons: LGRB sky positioning, weather, satellite location and localisation precision, to name but a few. Secondly, until recently, many follow-up programs were interested in very low and very high redshift GRBs. Such a biasing in combination with the redshift desert, results in a deficit, in the $z\sim1-5$ range, that is not easily removed by luminosity selections \citep[for an analysis of removing selection criteria see][]{Coward08}. These deficits can strongly bias the results and give different interpretations of metallicity cuts and evolving LGRB LFs (see Sect.~\ref{Results:OtherSamples}). Rather than improving completeness levels by cuts (e.g., luminosity, time criteria), Bayesian inference or other methods, incompleteness should be reduced by utilising programs with consistent LGRB follow-up. Such programs would require no selection biasing of GRB triggers for redshift or host follow-up. This is definitely not an easy aim and is not always possible with current telescope over-subscription, but many programs have already shown that completeness is an important criterion \citep[see e.g.,][]{Fynbo08, Greiner11, Kruehler11}.

In summary, the CSFH-LGRBR connection has a large number of free parameters that are co-dependent in several ways, especially if redshift dependencies are incorporated. As a result, each parameter cannot be treated independently and should always be considered contemporaneously, in a systematic way. Since the extrapolation of the individual parameters to higher redshifts is very difficult to determine, the most direct way of improving the uncertainties is primarily a systematic follow-up of all redshifts,  that would benefit from a GRB mission which has a substantially larger detection rate than {\it Swift} \citep[e.g., GRIPS;][]{Greiner11b}.
\section{Conclusion}
\label{sec:Conclusion}

The association of LGRBs with the death of massive stars has presented many new opportunities for utilising them as high redshift tools. One possibility is using the LGRB rate to trace the CSFH to unprecedented redshifts, which is usually challenging by conventional methods. To reach such a goal, the manner in which the LGRB rates traces the CSFH needs to be known, whether it is dependent on galaxy properties or not (e.g., galaxy stellar mass, metallicity or stellar initial mass functions). In this work we have studied whether any type of dependence is required, given new evidence of LGRBs occurring in more massive galaxies than had been previously thought in combination with highly complete redshift measured LGRB samples (up to 95\% complete). 

Using a highly complete sample we find best-fit solutions that show no preference for a strong metallicity or stellar mass constraints. These results imply that the LGRB population has no preference on the global properties of their host galaxy other than it has active star formation. We also show that our initial model does not require additional redshift dependences and, therefore, implying that there is no redshift dependence required in the LGRB probability or luminosity function.

The best-fit CSFH models are modified at redshifts $>3$ to linear functions to investigate the possibility that the CSFH flattens out. The least $\chi^{2}$ of the four modified models shows no preference for a flattening of star formation at high redshift. This is in contradiction with some other LGRB studies, but the exact form of the CSFH at high redshifts is still not settled. We predict that above $z=6$, $\sim1.2$\% of all LGRBs exist, which is in agreement with recent simulations \citep{Campisi10} and statistical studies \citep{Perley09, Fynbo09, Greiner11}.

Our results show that sample biasing and completeness levels of distributions are of essential importance and cannot always be recovered in the standard methods. Such completeness can only be achieved by consistent follow-up of LGRBs with no preference (or bias) on what LGRB is followed. Once unknown (and hard to quantify) biases are introduced, they can have dramatic changes to the interpretation of the data.

\begin{acknowledgements}
Part of the funding for GROND (both hardware as well as personnel) was generously granted from the Leibniz-Prize to Prof. G. Hasinger (DFG grant HA 1850/28-1). We thank Dominik Schleicher for discussions at the early stages of this project. We also thank T. Kr{\"u}hler, V. Sudilovsky, A. Rossi, D. A. Kann, M. Nardini, S. Klose, F. Olivares E. and S. Savaglio for constructive comments on the manuscript. JE thanks the support from B. Agarwal, G. Avvisati, C. F. P. Laporte, A. Longobardi and A. Monna. We thank the anonymous referee for the comments which improved the readability of the manuscript.
\end{acknowledgements}

\bibliographystyle{aa}
\bibliography{cosmology}

\end{document}